\documentclass[pra,amssymb,amsmath,superscriptaddress,twocolumn,footinbib,floatfix]{revtex4-2}
\usepackage{braket}
\usepackage{tikz}
\usepackage{multirow}
\usepackage{booktabs}
\usepackage[export]{adjustbox}
\usepackage{graphicx}
\usepackage{afterpage}
\usepackage[page,toc]{appendix}
\usepackage{amsthm}
\usepackage{xcolor}
\usepackage{dsfont}
\usepackage{hyperref}
\usepackage{chngcntr}
\hypersetup{
	colorlinks=true,
	linkcolor=blue,
	filecolor=magenta,      
	urlcolor=magenta,
	citecolor=teal
}

\usepackage[normalem]{ulem}
\usepackage{soul}
\DeclareMathOperator{\Tr}{Tr}

\newcommand{\outprod}[1]{\ket{#1}\!\!\bra{#1}}

\usepackage{dcolumn}
\usepackage{bm}
\usepackage{blkarray}
\usepackage{xparse}
\usepackage{mathtools}
\usepackage{subfiles}
\graphicspath{{figures/}}

\newcommand{\memk}[1]{\ket{\boldsymbol{#1}}}
\newcommand{\memb}[1]{\bra{\boldsymbol{#1}}}
\newcolumntype{P}[1]{>{\centering\arraybackslash}p{#1}}

\usepackage{float}

\begin{document}
\title{Entangling Quantum Memories via Heralded Photonic Bell Measurement}	
	\author{Prajit Dhara \thanks{\href{prajitd@arizona.edu}{prajitd@arizona.edu}}}%
 \email[]{prajitd@arizona.edu}
	\affiliation{Wyant College of Optical Sciences, The University of Arizona, Tucson, AZ 85721}
	\affiliation{NSF-ERC Center for Quantum Networks, The University of Arizona, Tucson, AZ 85721}
	\author{Dirk Englund}
 \email[]{englund@mit.edu}
	\affiliation{Research Laboratory of Electronics, Massachusetts Institute of Technology, Cambridge, MA 02139 }
		\affiliation{Department of Electrical Engineering and Computer Science, Massachusetts Institute of Technology, Cambridge, MA 02139 }
	\author{Saikat Guha}
 \email[]{saikat@arizona.edu}
	\affiliation{Wyant College of Optical Sciences, The University of Arizona, Tucson, AZ 85721}
	\affiliation{NSF-ERC Center for Quantum Networks, The University of Arizona, Tucson, AZ 85721}
 	\affiliation{Research Laboratory of Electronics, Massachusetts Institute of Technology, Cambridge, MA 02139 }
\begin{abstract}
		A common way to entangle quantum memories is via photonic entanglement swaps. Each of two memories, connected by an optical channel, emits a photonic qubit entangled with itself, and the photonic qubits undergo an entanglement swap on a beamsplitter in the middle of the channel. We compare two choices of encoding of the photonic qubit: single rail and dual rail. At low channel loss the dual-rail scheme outperforms the single rail scheme. However, as expected, the high-loss rate asymptote for the dual rail scheme scales quadratically worse with loss compared with single rail. Considering the following non-idealities: imperfect mode matching at the swap, carrier-phase mismatch across the interfered photonic qubits, and detector excess noise, we evaluate the density operator of the heralded two-qubit entangled state. We calculate a lower bound on its distillable entanglement per copy, and its Fidelity (with the ideal Bell state). For both schemes, imperfect swap-visibility results in a constant-factor decrease in the rate, while excess noise results in a dropoff of distillable entanglement beyond a certain total channel loss threshold, to zero. Despite the single-rail scheme's better rate-loss scaling, it is more severely affected by excess noise. The single-rail scheme is adversely affected by stochastic carrier-phase mismatch, which does not affect the dual-rail scheme. We study entanglement distillation on the heralded noisy entangled states for both methods, and outline a suite of quantum networking studies that our work could incite. 
	\end{abstract}
	
	\maketitle
	\section{Introduction}
	The most common way to generate entanglement between a pair of `matter' qubits---be it trapped-ions, solid-state defect centers, neutral atoms, or superconducting qubits---is by first generating photons entangled with each qubit, and performing a Bell State Measurement (BSM) on the photonic qubits via a beamsplitter and photon detectors, which succeeds with some probability. The probability of success drops with the overall loss in the photons' lifetimes, from the time they were generated (entangled with the matter qubit) to when they were detected. When the BSM fails, the matter qubits are re-initialized and re-used. But when the BSM succeeds, the two matter qubits are `heralded' in a two-qubit entangled state, whose fidelity can be quite high, especially if the matter qubits can be held with negligible drop in fidelity up until the success-failure information about the BSM arrives back at the matter-memory sites. This method has been proposed and implemented for the generation of entanglement between solid state spin qubits~\cite{Barrett2004-sj,Kalb2017-hr,Pompili2021-bp,Hermans2022-zl,Gao2012-eh}, trapped ion qubits~\cite{Hucul2014-jj, Inlek2017-tg, Stephenson2020-rw, Krutyanskiy2022-vh,Bock2018-qx,Dudin2010-ip}, neutral atoms~\cite{Van_Leent2020-ga,Van_Leent2022-yz,Ikuta2018-tb,Borregaard2015-uo} and superconducting qubits~\cite{Magnard2020-dc,Krastanov2021-pq}.
 
	
	Compared to the \emph{source in the middle}~\cite{Jones2013-wh,Jones2016-mg} method for entanglement generation, in the present approach, matter quantum memories themselves are the source of the entangling photons. This eliminates the complexity of building a reliable high rate high-fidelity entangled pair source and field-deploying the same, at the potential cost of stricter network operational requirements. Previous works examining the linear optics based entanglement swap protocol have covered various crucial aspect of the topic. Entanglement swapping between nitrogen-vacancy color center spin qubits in diamond is analyzed in~\cite{Hermans2023-kf,Goodenough2021-nt} with specific focus on the novelties and challenges of solid state spin qubits. An alternative approach to the quantum state analysis has been covered in~\cite{Wein2020-vp}, with focus on the time dynamics of the optical entanglement swap for various photonic encoding choices. 
	
	Most studies of entanglement distribution protocols and platforms utilize the quantum state fidelity (with ideal QM Bell pairs) as the decisive and quantitative `metric' \footnote{The word metric is used here without any mathematical rigor.} for the protocol performance. The difficulty arises in the fact that sub-unity fidelity has different implications on the utility of the distributed entangled state. The current article utilizes the hashing bound of the modelled state, which serves as a lower bound to the state's distillable entanglement. 
	
	 In this work, we derive and analyze the density matrix description of two quantum memories entangled using a photonic Bell state measurement. Our analysis and derivation takes into account the various non-idealities introduced in the entanglement swapping i.e.,  loss in photonic qubits, imperfect detection efficiency, excess noise in the channel, photonic mode mismatch (i.e., visibility of the interference at the beamsplitter)and carrier-level phase mismatch. Our analysis of the protocol is done using the hashing bound; we highlight the key differences in the performance of the encoding choices. Further, most existing studies use the Werner state model of two-qubit entanglement for studying entanglement distillation. Our analysis shows that neither photonic qubit encodings described above result in a Werner state heralded among the memory qubits. We study the performance of simple entanglement distillation schemes on the true heralded entangled states. 
 
    The article is organized as follows. The swap setup and associated definitions are covered in Sec.~\ref{sec:sys_conds}. Section~\ref{sec:ent_swap_analysis} analyzes the states generated by  ideal entanglement swaps and presents the fundamental tradeoffs. A detailed analysis of the entanglement swapping with non-idealities is covered in Sec.~\ref{sec:tradeoffs}. In Sec.~\ref{sec:distillation}, we analyze the effect of quantum state distillation, and highlight the link level improvements in the heralded quantum state. Section~\ref{sec:conclusion} concludes the analysis with discussion of potential applications for the underlying models and proposals for improved swapping.
	
	\section{System Considerations}
	\label{sec:sys_conds}
The setup analyzed in this article is a generalization of entanglement swapping between a wide class of physical qubit implementations. We consider two parties, Alice $ (A) $ and Bob $ (B) $ equipped with \emph{emissive quantum memories} (QM). For the entirety of this manuscript, emissive QMs denote systems which are able to generate an entangled state of the internal state of the QM and one or more photonic modes.

The emitted optical modes are transmitted over a loss-prone transmission media (for e.g. optical fiber, dielectric waveguides, or free-space optical links) to a central node Charlie $ (C) $, which performs a linear optical entanglement swap. The actual implementation of the swap is encoding choice dependent; however, the swap success outcome (where the swaps heralds the generation of shared entanglement between $ A $ and $ B $) is probabilistic. The optical channels between $ A-C $ and $ B-C $ are modeled as pure loss bosonic channels, with associated transmissivities $ \eta_A $ and $ \eta_B $ respectively, where $ \eta_k \in[0,1]$, which maybe expressed in dB as $ \eta_{\mathrm{dB}}=-10 \log_{10} \eta $. Readers should note that any additional losses (say in the memory to photon interface or in the detectors) can be lumped into the channel transmissivity parameters. The actual physical details of the quantum channel may be wide ranging --- our analysis holds true for communication on photonic waveguides networking solid state memories on a chip ($ \eta_{\mathrm{dB}}\sim0-3 $ dB), optical fiber links at the metropolitan scale ($ \eta_{\mathrm{dB}}\sim1-10 $ dB) or free-space optical links for satellite based communications ($ \eta_{\mathrm{dB}}\sim10-10^2 $ dB).


\begin{figure}[ht!]
	\centering
	\includegraphics[width=\linewidth]{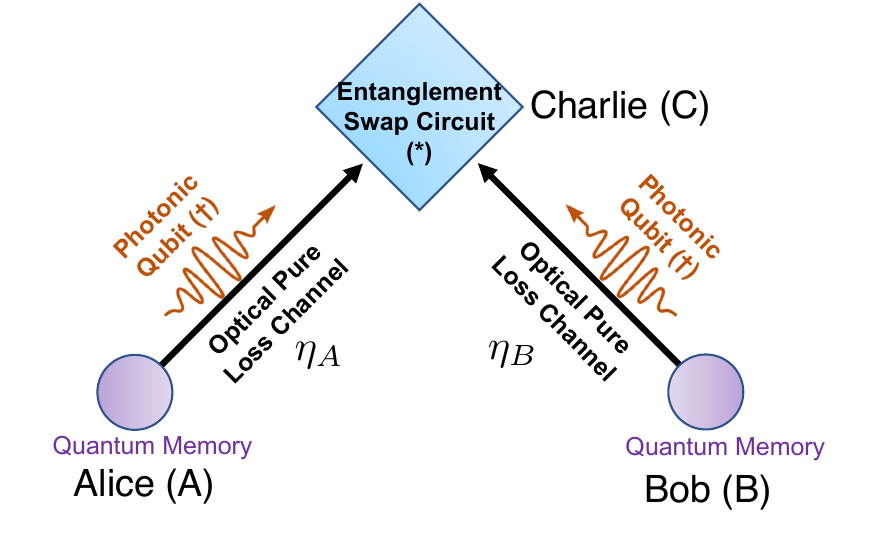}
	\caption{Midpoint entanglement swap for photonic qubits (orange) entangled to matter qubits (purple). The entanglement swapping circuit (blue diamond) is photonic qubit encoding dependent.  Appendix~\ref{app:general_state} covers detailed descriptions for the swaps.}
	\label{fig:compveta}
\end{figure}

The final degree of freedom in this system is the choice of photonic qubit encoding format. We limit our analysis and discussion to discrete variable encoding formats which rely on the vacuum and single photon states of an bosonic mode. The \emph{single rail} encoding, is defined by the presence or absence of a photon in a single bosonic mode; the logical qubit states for the mode labeled by $ k $ are defined by 
\begin{align}
	\ket{\bar{0}} \equiv \ket{0}_k ;\ket{\bar{1}}\equiv \ket{1}_k.  
\end{align}
An alternative is the \emph{dual rail} encoding, the logical qubit states are represented by the presence of a single photon in one of two orthogonal bosonic modes labeled by $ k1 $ and $ k2 $ (which may be spatial, temporal, spectral modes or any combination thereof), as
\begin{align}
	\begin{split}
		\ket{\bar{0}}\equiv\ket{1}_{k1}\ket{0}_{k2} =\ket{1,0}_{k};\\
		\ket{\bar{1}}\equiv\ket{0}_{k1}\ket{1}_{k2}=\ket{0,1}_k .
	\end{split}
\end{align}

Before we proceed with the analysis of the entanglement swapping circuitry, additional notation for the local entangled pair (namely between the quantum memory and the photonic qubit) has to be established. We consider general entangled states of the form,
\begin{align}
	\ket{\psi (\gamma)}_{S}=\sqrt{\gamma}\memk{1}_S \ket{\bar{0}}_{S,k}+\sqrt{1-\gamma}\memk{0}_S \ket{\bar{1}}_{S,k}.
	\label{eq:init_epair}
\end{align}
where $ \memk{0} $ and $ \memk{1} $ represent the qubit levels of the QM and the subscript $ S =\{A,B\}$ is used to denote the party that posses the corresponding memory. The generation of the state in Eq.~\eqref{eq:init_epair} is again qubit hardware and photonic qubit encoding dependent~\cite{Barrett2004-sj,Pompili2021-bp,Hermans2022-zl,Gao2012-eh,Hucul2014-jj, Inlek2017-tg, Stephenson2020-rw, Krutyanskiy2022-vh,Magnard2020-dc,Krastanov2021-pq}. Fig.~\ref{fig:compveta} depicts this abstractly, with purple `matter' qubits and orange photonic qubits. The photonic qubits are transmitted over an optical channel or link before they meet at the entanglement swapping circuit (blue diamond). 

The details of the swap are encoding dependent and are discussed in Appendix~\ref{app:general_state} (see Fig.~\ref{fig:dr_swaps}). In the most general scenario, the photonic qubits are mixed on a balanced (50:50) beamsplitter (for erasing the which source/path information) and detected by photon number resolving detectors. Detection of specific click patterns heralds the successful generation of an entangled state (say $ {\rho}_{AB} $) shared between the QMs along with the entangled state parity information. Since swaps are probabilistic, we denote the probability of successfully heralding the state by $ P_{\mathrm{succ.}} $.

Of the most common metrics to evaluate the `quality of entanglement', \emph{state fidelity} evaluates the  overlap of the final state $ {\rho}_{AB} $ with the ideal target state (usually a Bell state, here we choose $ \ket{\Psi^+} = \left(\memk{0}_A\memk{1}_B \pm \memk{1}_A\memk{0}_B\right)/{\sqrt{2}}$). The state fidelity is evaluated by 
\begin{align}
	F({\rho}_{AB},\ket{\Psi^+})\coloneqq\braket{\Psi^+|{\rho}_{AB}|\Psi^+}.
\end{align} 
Fidelity is a reliable and insightful state quality indicator, but only in the regime where $ F(\cdot) $ is close to unity. In the context of shared entanglement generation, evaluating (or bounding) the \textit{distillable entanglement} of the final state is more insightful. Distillable entanglement, represented by $ E_D(\rho_{AB}) $, quantifies the number of perfect entangled pairs (Bell pairs) that can be distilled from $ \rho_{AB} $, assuming both parties have \emph{ideal universal quantum computers } (using an arbitrary non-specified distillation circuit) and unlimited two-way classical communications. For general states, $ E_D(\rho_{AB}) $ is non-trivial to evaluate; for the present study we will use the  \textit{hashing bound} $ I(\rho_{AB})$, which is a lower bound to the state's distillable entanglement. The hashing bound is 
\begin{align}
	I(\rho_{AB})=\max[S(\rho_{A})-S(\rho_{AB}),S(\rho_{B})-S(\rho_{AB})],
\end{align}
where, $ \rho_A =\Tr _B(\rho_{AB})$, $ \rho_B =\Tr _A(\rho_{AB})$ and $ S(\rho)$ is the von-Neumann entropy of the state $ \rho $. 

The product of the hashing bound  (units: ebits per state copy) and the probability of success (units: states per swap attempt) yields a lower bound to the distillable entanglement rate. Henceforth we will use the symbol $ \mathcal{R}(\rho) $ to denote this where,
\begin{align}
	\mathcal{R}(\rho) =	I(\rho)\times P_{\mathrm{succ.}}.
	\label{eqn:ult_rate}
\end{align} 
For any protocol/encoding choice the rates are upper bounded by the repeater-less bound to the channel capacity, given by $ D_2(\eta)= -\log_2(1-\sqrt{\eta}) \approx 1.44 \sqrt{\eta}  \text{ for }\eta\ll1$~\cite{Pirandola2017-qr}. { $D_2(\eta)$ specifies the ultimate limit to the entanglement generation rate (in ebits per channel use) for two parties communicating over a pure loss channel of transmissivity $\eta$.} Thus, we expect $ I(\rho_{AB}) \leq E_D(\rho_{AB})<D_2(\eta)$ for all values of $ \eta $.

	\section{Ideal Entanglement Swapping}
	\label{sec:ent_swap_analysis}
With ideal hardware, number resolving detectors, and no optical transmission loss $ (\eta_A=\eta_B=1) $ the linear optical entanglement swap succeeds with a maximal $ P_{\mathrm{succ.}}=1/2$~\cite{Calsamiglia2001-vp}, irrespective of the choice of encoding. However, in the presence of loss, the action of the swap is quite distinct between the two encoding formats. To simplify analysis, we consider the  specific case of symmetric channel losses i.e.\ $\eta_A=\eta_B=\sqrt{\eta} $ for a total channel transmissivity of $ \eta_A\eta_B =\eta$. 

When mode $ k $ of the state in Eq.~\eqref{eq:init_epair} is dual rail encoded, the final state is heralded with the probability of success  $ P_{\mathrm{succ.}}= 2\gamma\,(1-\gamma)\eta $, and {  is described by},
\begin{align}
	{\rho}_{\mathrm{dual}}=\frac{1}{2}\left(\memk{0}_A\memk{1}_B \pm \memk{1}_A\memk{0}_B\right) \left(\memb{0}_A\memb{1}_B\pm\memb{1}_A\memb{0}_B\right).  
\end{align}
{  It is straightforward to note that the density operator }$ {\rho}_{\mathrm{dual}} $ has unit fidelity with the corresponding QM Bell states  $ \ket{\Psi^\pm}= \left(\memk{0}_A\memk{1}_B \pm \memk{1}_A\memk{0}_B\right)/{\sqrt{2}}$, implying that pure loss causes no detriment to the quality of the state. Additionally, the optimal value of $ \gamma $ that maximizes $ P_{\mathrm{succ.}} $  at a given value of $ \eta $ is \textit{always} $ 1/2 $.


In contrast, when the single rail encoding is used, the final state is heralded with $ P_{\mathrm{succ.}}=2\sqrt{\eta}\,(1-\gamma)(1-(1-\gamma)\sqrt{\eta})$ and with the final state density operator,
\begin{align}
	\begin{split}
		{\rho}_{\mathrm{single}}=&\frac{\alpha_1}{2}\!\cdot \left(\memk{0}_A\!\memk{1}_B \pm \memk{1}_A\!\memk{0}_B\right) \left(\memb{0}_A\!\memb{1}_B\pm\memb{1}_A\!\memb{0}_B\right) \\
		&+ \alpha_2 \cdot\memk{0}\!\!\memb{0}_A\otimes \memk{0}\!\!\memb{0}_B,
	\end{split}
\end{align}
whose coefficients $ \alpha_k $ are given in terms of the parameters as,
\begin{subequations}
	\begin{align}
		\alpha_1&=\gamma \,(1 - \gamma) \,\sqrt{\eta} /P_{\mathrm{succ.}}, \\
		\alpha_2&= \sqrt{\eta} (1-\gamma)^2\left(1-\sqrt{\eta}\right)/ P_{\mathrm{succ.}}.
	\end{align}
\end{subequations}
Readers may note that when $ \eta<1 $, the state fidelity of $ \rho_{\mathrm{single}} $ w.r.t.\ the ideal Bell pair $ \ket{\Psi^+}\equiv(\ket{0,1}+\ket{1,0})/\sqrt{2} $ is always less than one, since $ \alpha_2\neq0\, \forall \, \eta\in\left(0,1\right)$ . The $ \alpha_2 $ term only becomes zero for either $ \eta=1 $ (perfect transmission), $ \eta=0 $ (no transmission) or when $ \gamma=1 $ (initial state is no longer entangled). One can thus evaluate an optimal value of $ \gamma $ which maximizes $ P_{\mathrm{succ.}}$ with a constraint on some other heralded state metric (such as fidelity, or distillable entanglement).

	\section{Tradeoffs for Non-Ideal Swaps}
	\label{sec:tradeoffs}

\subsection{Problem Setup}

We extend the analysis of the previous section by considering hardware and channel non-idealities which would be relevant for any practical entanglement swapping link - (1) excess noise in channel/detectors, (2) imperfect mode matching and (3) carrier level phase mismatch of the bosonic modes. The complete state descriptions with these detriments accounted for is derived in Appendix~\ref{app:general_state}. We present the state descriptions in the logical and Bell basis, along with state metric evaluation for a few corner cases and results for asymmetric links in Appendix~\ref{app:detailed_state_deriv}. Our methodology for modeling these non-idealities are summarized below.

\emph{Excess noise} --- Excess noise in the system can arise from background photons in the channel, electronic Johnson-Nyquist noise in detectors and detector dark clicks. All of these effects can be lumped into a single parameter, namely, the excess photons per mode, which we shall denote using the symbol $ P_d $ with a typical range $ P_d\in\left[0,1\right) $. For a non-zero $ P_d $ the final quantum state becomes mixed and has additional terms whose proportion in the total state are of the {  order of $P_d$, or higher}.
{ The excess photon rate can be determined as a product of $P_d$ (units: excess photons per mode) and inverse of the temporal length of a photonic mode (units: modes per second). }

\emph{Imperfect mode matching } --- Mode matching is crucial for the photonic entanglement swap to function properly, as  any analysis of the modes post-interference must not reveal the path or `which memory' information. Imperfect mode matching leads to inherent distinguishability  of the interacting photons, which is detrimental to enatnaglement swapping. The effect of mode mismatch is compactly described by a visibility parameter $ \mathcal{V} \in[0,1]$, where $ \mathcal{V}=1 $ denotes perfect mode matching. This one number $\mathcal{V}$ can account for the total mode mismatch in the spatio-temporal-polarization mode of the photonic-qubit-bearing mode pairs emanating from the two memory sites, being interfered on the swapping beamsplitter.


\emph{Carrier Phase Mismatch} --- In addition to mode matching, any imprecision in the optical carrier phase matching is an important parameter for entanglement swapping. This is especially relevant for the single-rail scheme as the memory-qubit photonic-qubit entangled states from the two sides undergo a fast-oscillating carrier phase based on their total propagation length. This fast-oscillating carrier phase is much more difficult to lock compared to the spatio-temporal modes of the photonic modes, arriving from the two memory sites. The carrier phase is imparted to the system by applying the unitary $ U(\theta)=\exp(i\theta \hat{n}) $ to the bosonic mode(s), where $ \hat{n} $ is the modal number operator and $\theta$ is a random phase drawn from a zero-mean normal distribution with variance $ \varepsilon $, i.e., $\theta\sim \mathcal{N}(0,\varepsilon) $. Phase mismatch is treated by introducing it as a complex visibility parameter which may be combined with the mode mismatch parameter to yield a single complex quantity $\mathcal{V}=|\mathcal{V}|e^{i\theta}$ with  $ |\mathcal{V}|\in[0,1] $ and $ \theta\sim \mathcal{N}(0,\varepsilon) $. The stochastic nature of $ \theta $ necessitates evaluating the ensemble averaged quantum state of the two heralded memory qubits, to examine the effect of phase mismatch. { Typically $\varepsilon\ll 2\pi$; however for large phase variance, it becomes a uniform random variable over the range $[-\pi,\pi)$. When transmission from $A-C$ and/or $B-C$ is done over a free space optical link (for e.g., between two terrestrial sites for $A,B$ and a space-borne/aerial platform for $C$), the phase variance is affected by atmospheric turbulence; it must be corrected for using phase-compensation techniques. Appendix~\ref{app:phase_mismatch} reports expected phase variances in such implementations. }

\begin{figure}[htbp]
	\centering
	\includegraphics[width=\linewidth]{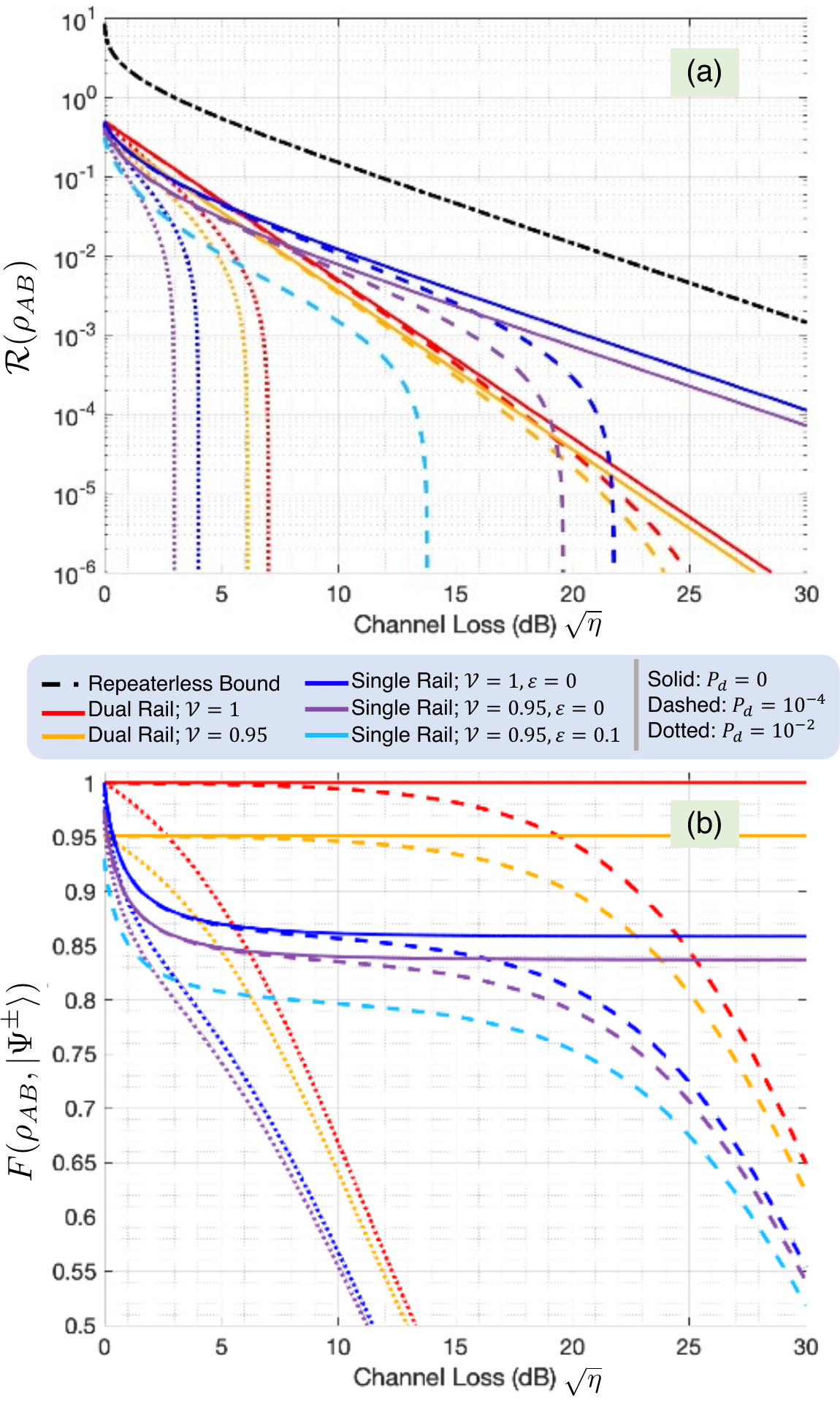}
	\caption{Comparison of the single rail swap (blue for $ |\mathcal{V}|=1,\varepsilon=0 $, purple for $ \mathcal{V}=0.95,\varepsilon=0 $, cyan for $ \mathcal{V}=0.95,\varepsilon=0.1 $) with the dual rail swap (red for $ |\mathcal{V}|=1$, orange for $ \mathcal{V}=0.95$) using the (a) hashing bound rate $\mathcal{R}(\rho_{AB})$ where we compare them to repeater less bound (black) and, (b) state fidelity $F(\rho_{AB},\ket{\Psi^\pm})$ .The effect of non-zero excess noise is highlighted by the various line styles: solid for $ P_d=0 $, dashed for $ P_d=10^{-4} $ and dotted for $ P_d=10^{-2} $.  }
	\label{fig:tradeoffs}
\end{figure}

\subsection{Tradeoffs}
We examine the trade-off between the different encoding choices for our entanglement generation link, by evaluating the hashing bound rate $ \mathcal{R}(\rho_{AB}) $ and state fidelity, $ F(\rho_{AB},\ket{\Psi^\pm}) $ in Fig.~\ref{fig:tradeoffs}(a) and (b) respectively. We examine and compare swaps employing dual rail encoding (depicted by the red for $|\mathcal{V}|=1 $ and orange for $ |\mathcal{V}|=0.95 $) and the single rail encoding (blue for $ |\mathcal{V}|=1;\varepsilon=0 $, purple for $ |\mathcal{V}|=0.95;\varepsilon=0 $ and cyan for $ |\mathcal{V}|=0.95;\varepsilon=0.1 $). For the single rail swap, the qubit parameter $ \gamma $ is chosen to maximize $ \mathcal{R}(\rho_{AB}) $ at a given $ \eta $. The black dot-dashed curve in Fig.~\ref{fig:tradeoffs}(a) is the repeater-less bound of $ D_2(\eta) $~\cite{Pirandola2017-qr}. For all encodings, the line style depicts the excess noise in the channel: solid lines for $ P_d=0 $, dashed lines for $ P_d=10^{-4} $ and  dotted lines for  $P_d=10^{-2}$). We make the following observations ---
\begin{itemize}
	\item In terms of {  distillable entanglement swapping rate} $ \mathcal{R}(\rho_{AB}) $, the single rail { encoding} outperforms the dual rail { encoding} (with a scaling of $ \mathcal{O}(\eta) $) in the high loss limit $ (\eta\ll1) $. However, in the low loss limit ($ \eta <16 $ dB), the dual rail { encoding} outperforms the single rail { encoding}.
	
	\item Swapping with $ |\mathcal{V}|<1$, does not affect $ P_{\mathrm{succ.}} $ but affects the  state quality  $ I(\rho_{AB}) $, which is lower than the ideal unity visibility state for all  $ \eta$.

	\item Phase mismatch only affects the single rail state. The visibility parameter of the ensemble averaged state  is modified as $ \mathcal{V}\rightarrow\mathcal{V}\times \exp(-\varepsilon)\approx\mathcal{V}(1-\varepsilon); \text{ when }\varepsilon $ is small. We see the corresponding effect of a lowered $ \mathcal{V} $ on $ \mathcal{R}(\rho_{AB}) $ and $ F(\rho_{AB},\ket{\Psi^\pm}) $ (cyan curve). 
	
	\item For $ P_d>0 $, $ \mathcal{R}(\rho_{AB}) $ crashes to zero at a finite value of $ \eta $, indicating a maximum range for entanglement swapping in the presence of excess noise. However, $ F(\rho_{AB},\ket{\Psi^\pm}) $ doesn't collapse to a value less than $0.5$ for the corresponding states. Additionally, the single rail swapped state is more susceptible to excess noise, i.e., for same $P_d$, $ \mathcal{R}(\rho_{AB}) $ crashes to zero for lower loss for the single rail swapped state.
	
	\item In the presence of excess noise and mode mismatch, states (heralded by either single or dual rail swaps) with lower $ |\mathcal{V}| $ are more susceptible to loss.
	
\end{itemize}
The difference in the rate scaling is quite evident from the expressions of $P_{\mathrm{succ.}} $ as highlighted in Section~\ref{sec:ent_swap_analysis}. For $ \eta\ll1 $, upto the highest order the single rail swap is $ \mathcal{O}(\sqrt{\eta}) $ and dual rail swap is  $ \mathcal{O}({\eta}) $. Intuitively, this makes sense as well, the single rail swap requires the successful transmission of a single photon to the midpoint over half of the entire link $(\propto\sqrt{\eta} )$, whereas the dual rail swap requires two photons (one from A and B each).

\begin{figure}[htbp]
	\centering
	\includegraphics[width=\linewidth]{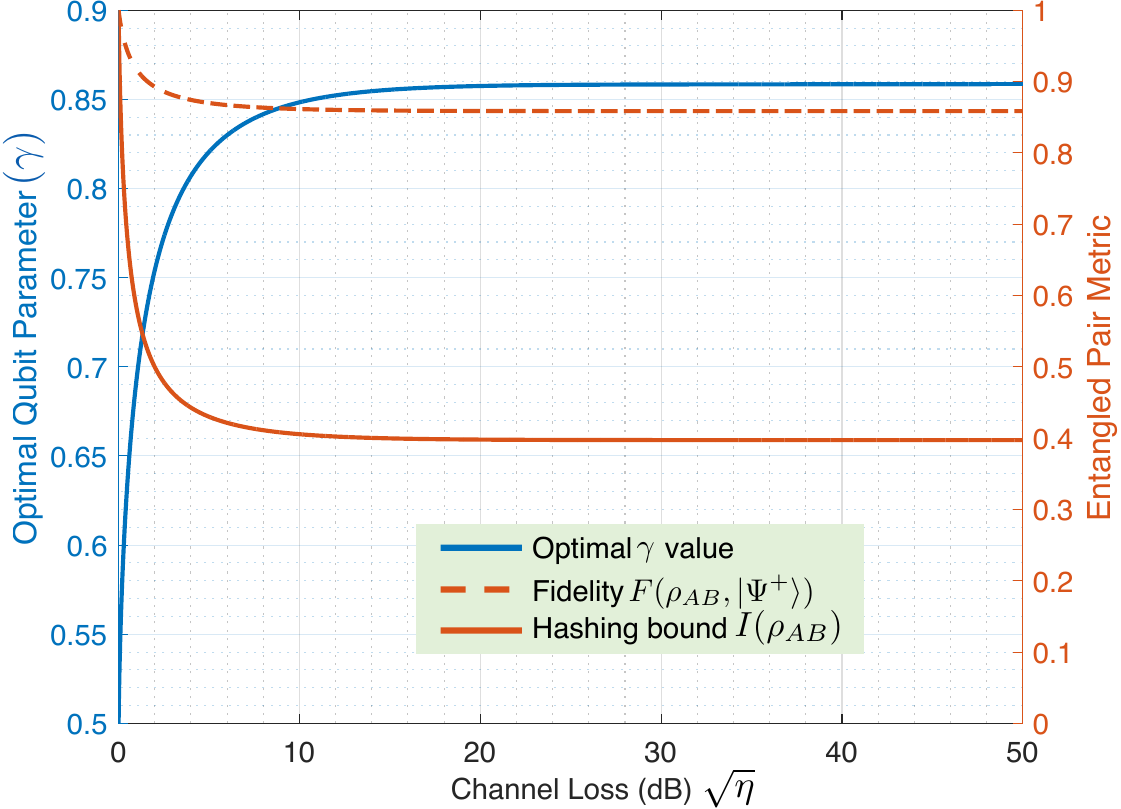}
	\caption{State fidelity $ F(\rho_{AB},\ket{\Psi^\pm}) $ (orange solid) and hashing bound $ I(\rho_{AB}) $ (orange dashed) for the single-rail entangelment swap, plotted with optimal qubit initialization parameter $ \gamma $ (blue) for different values of channel loss. We assume $P_d=0, \mathcal{V}=1,\varepsilon=0$ for this plot. }
	\label{fig:opti_gamma}
\end{figure}

The non-trivial behavior of the single rail swap rate scaling for $ \eta\rightarrow1 $ regime arises as a consequence of the optimization of $ \gamma $ to maximize $ \mathcal{R}(\rho_{AB}) $. Consider the ideal swap with loss i.e., $ |\mathcal{V}|=1 $ and $ P_d=0 $, for which the expressions for successful swap probability and state hashing bound are respectively given as,
\begin{subequations}
	\begin{align}
		P_{\mathrm{succ.}}&=2\sqrt{\eta}(1-\gamma)(1-(1-\gamma)\sqrt{\eta}); \label{eq:ps_single_sym}\\
		I(\rho)&=h_2\biggl(\frac{\gamma/2}{1-\sqrt{\eta}(1-\gamma)}\biggr)-h_2\biggl(\frac{\gamma}{1-\sqrt{\eta}(1-\gamma)}\biggr)
	\end{align}
\end{subequations}
where $ h_2(x)=-x \log_2 x-(1-x) \log_2(1-x) $ is the binary entropy function. Numerical maximization of $ \mathcal{R}(\rho_{AB})$ yields the optimal $ \gamma $ as function of $ \eta $ as depicted in Fig.~\ref{fig:opti_gamma} (blue line). We plot the state fidelity (orange dashed) and distillable entanglement (orange solid) for the corresponding values of $ \eta $. For $ \eta=1 $, $ I(\rho_{AB}) $ attains the maximal value of 1 ebit per copy. For this instance, $ P_{\mathrm{succ.}} $ becomes $ 2\gamma(1-\gamma) $ which is maximized for $ \gamma=1/2 $. In the $ \eta\ll1 $ regime, $ P_{\mathrm{succ.}}\approx2\sqrt{\eta}(1-\gamma) $ and correspondingly $ I(\rho)\approx h_2(\gamma/2)-h_2(\gamma) $. Here $ \mathcal{R}(\rho) $ has $ \sqrt{\eta} $ as a pure multiplicative factor; this means the optimal $ \gamma $ that maximizes it is independent of $ \eta $ and is a solution of the transcendental equation,
\begin{align}
	\frac{\partial}{\partial \gamma}\left[(1-\gamma)(h_2(\gamma/2)-h_2(\gamma))\right]=0.
\end{align}

{  The effect of imperfect mode-matching manifests as a sub-unity swap visibility, i.e. $ |\mathcal{V}|<1 $. This effect can be analyzed by expressing the final state in the Bell basis. As an example, the reader may consider the noiseless symmetric dual rail swapped state ($ P_d=0; \eta_A=\eta_B=\sqrt{\eta}, \eta_d=1 $; see Appendix~\ref{app:subsec_dual_deriv} for the detailed derivation and Appendix~\ref{app:subsec_dual_desc} for state description), which may be expressed in the Bell basis as, 	
\begin{align}
	{\rho}_{AB}=\frac{({1+\mathcal{V}^2})}{2} \outprod{\Psi^+ }+\frac{({1-\mathcal{V}^2})}{2} \outprod{\Psi^- }.
\end{align}
whose $ I(\rho_{AB})=1-h_2((1-\mathcal{V}^2)/2)$. This quantity decreases monotonically with $ \mathcal{V} $. The corresponding probablity of successfully heralding an entangled state is  $P_{\mathrm{succ.}}=\eta/2$, which is independent of $\mathcal{V}$. Hence in Fig.~\ref{fig:tradeoffs} the $ \mathcal{R}(\rho_{AB}) $ curve for $ |\mathcal{V}|=0.95 $ is strictly below the curve for $ |\mathcal{V}|=1 $. A similar analysis can be performed for the single rail state; for the noiseless symmetric single rail swapped state the probability of success is given in Eq.~\eqref{eq:ps_single_sym} which is independent of $\mathcal{V}$. The expression for $I(\rho_{AB})$ involves $\mathcal{V}$; however the analytic form of the expression does not give us any intuitive insights and we have to resort to numerical evaluation to show $I(\rho_{AB})$ decreases with $\mathcal{V}$  (see Appendix~\ref{app:subsec_single_deriv} for the detailed derivation; Appendix~\ref{app:subsec_single_desc} for state description).}

{
	The effect of phase mismatch is an important distinction between two encodings. The detailed state description and derivation (see Appendix~\ref{app:detailed_state_deriv}-\ref{app:general_state})
	highlight this difference in how phase mismatch affects the initial quantum state~\cite{Barrett2004-sj}. The initial single rail memory-photon state has the form 
	\begin{align}
		\ket{\psi}_{\mathrm{mem,photon}}= \sqrt{\gamma}\memk{1}_S\ket{0}_{S,k} +e^{i\theta_S} \sqrt{1-\gamma}\memk{0}_{S}\ket{1}_{S,k},
	\end{align}
	 whereas the optimal initial state for the dual rail swap is expressible as 
	\begin{align}
		\ket{\psi}_{\mathrm{mem,photon}}= e^{i\theta_S} [\memk{1}_S\ket{1,0}_{S,k} + \memk{0}_{S}\ket{0,1}_{S,k}].
	\end{align}
	The phase mismatch effect for the former hence introduces a relative phase, whereas it manifests as global phase for the latter. Consequently upon entanglement swapping of the photonic qubits, the off-diagonal elements $\braket{\mathbf{0,1}|\rho_{\mathrm{single}}|\mathbf{1,0}}$ or $\braket{\mathbf{1,0}|\rho_{\mathrm{single}}|\mathbf{0,1}}$, of the single rail heralded state is dependent on a complex $\mathcal{V}$ term where $ \arg \mathcal{V}=\theta_A-\theta_B $.  On the other hand, the dual rail state is only dependent on the  magnitude, i.e.\ $|\mathcal{V}|$. 
	
	Readers should note that the carrier phase difference $ \arg \mathcal{V}$ is a stochastic parameter. To characterize the state performance, it is necessary to evaluate the ensemble averaged state. Since we typically consider $ \theta_A,\theta_B\sim\mathcal{N}(0,\varepsilon) $, this implies $\arg \mathcal{V} \sim\mathcal{N}(0,2\varepsilon) $. Under these considerations, the ensemble averaged state's visibility is modified as $ \mathcal{V}\rightarrow \mathcal{V}\exp(-\varepsilon)\approx\mathcal{V}(1-\varepsilon)$ for small $\varepsilon  $ (see derivation in Appendix).
}

\begin{figure}[h!]
	\centering
	\includegraphics[width=\linewidth]{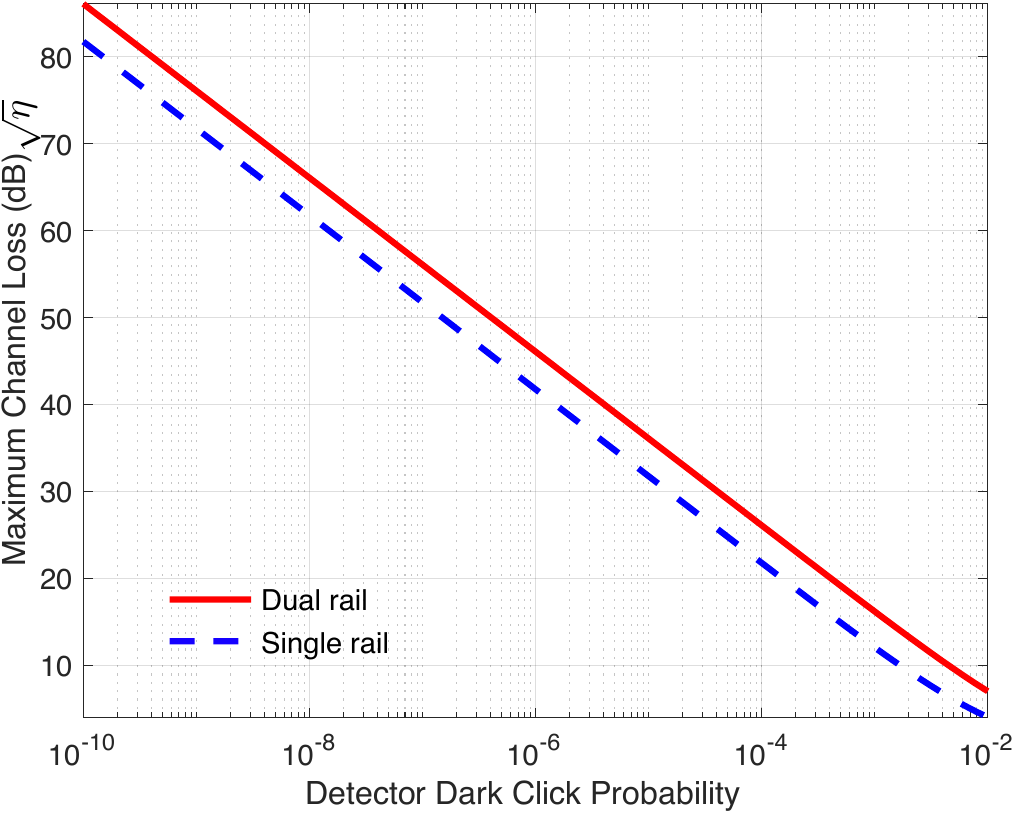}
	\caption{Maximum range of heralded states by dual (red) and single (blue dashed) rail swaps for specified values of excess noise $ P_d $. Assume $\mathcal{V}=1,\varepsilon=0$.}
	\label{fig:dark_clk}
\end{figure}

The effect of excess noise on the state is evident when we consider the detailed state description. As an example, consider the heralded dual rail state for the symmetric case ($ \eta_A=\eta_B=\sqrt{\eta},\, \eta_d=1 $ and $ \mathcal{V}=1 $; details in Appendix~\ref{app:general_state}). The complete state may be expressed in terms of the Bell states $ \{\ket{\Psi^\pm}, \ket{\Phi^\pm}\} $ as,
\begin{align}
	\rho_{AB}=& \beta_1 \outprod{\Psi^+}+\beta_2 \mathbb{I}_4,
\end{align}
where $ \mathbb{I}_4 $ is the 4-dimensional identity operator and,
\begin{subequations}
	{\small
		\begin{align}
			\beta_1&=\frac{(1-P_d)^4 \times \eta }{2\mathbf{N}_d},\\
			\beta_2&=\frac{P_d(1-P_d)^2}{\mathbf{N}_d}\biggl(\frac{1}{2}(1-P_d)\sqrt{\eta}(1-\sqrt{\eta})+P_d(1-\sqrt{\eta})^2\biggr),
	\end{align}} 
\end{subequations}
with $ \mathbf{N}_d$ ensuring $ \Tr(\rho_{AB}) =1$ .
It is clear that {  the state component} that contributes to `usable' entanglement is the $ \outprod{\Psi^+} $ component. The complementary terms is the maximally mixed two qubit state, and yields no distillable entanglement. Hence for all $ P_d >0$, the contribution of $ \beta_2 $ would surpass that of $ \beta_1 $ at some finite value of $ \eta $ which limits $ \mathcal{R}(\rho_{AB}) $ by setting $ I(\rho_{AB}) =0$. We call this the \emph{maximum range } of the protocol, since the state descriptions indicates minimal usable entanglement beyond this value of $ \eta $. General formulae can be obtained by equating $ I(\rho_{AB}) =0$; however very minimum insight is gained for the true maximum range trend. In lieu of analytically expressible values, Fig.~\ref{fig:dark_clk} plots the numerically extracted value of the maximum range for various values of $ P_d $.  

\begin{figure}[h!]
	\centering
	\includegraphics[width=\linewidth]{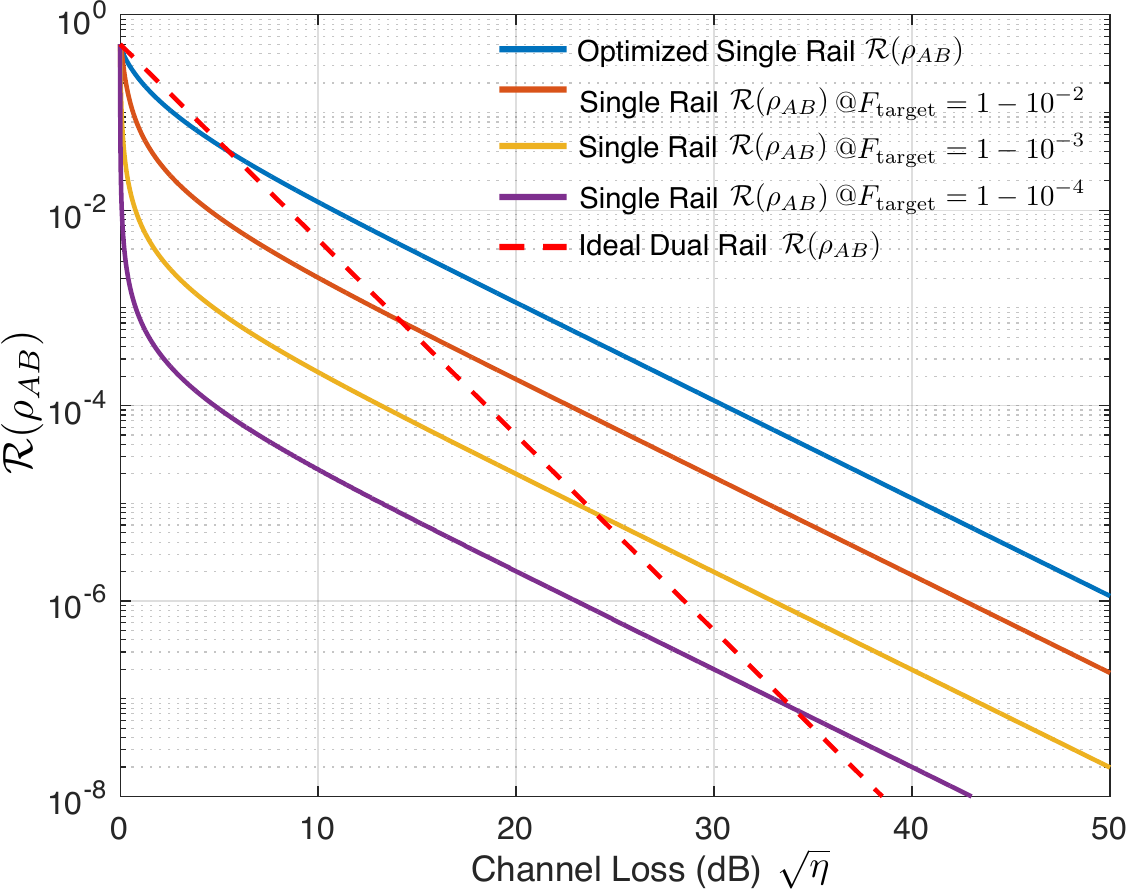}
	\caption{Distillable entanglement rate for the single rail swap at target fidelities (orange: $ 1-10^{-2} $, yellow: $ 1-10^{-3} $, purple: $ 1-10^{-4} $) compared to optimal rate ($ F\approx0.858$) for $\mathcal{V}=1,\varepsilon=0$.. The ideal dual rail swap rate (at $ \mathcal{V}=1; P_d =0$; red dashed line) is given for comparison.}
	\label{fig:rate_targFid}
\end{figure}

For the single rail swap, readers may also find it interesting to note that the optimal value of $ \gamma $ yields a state with fidelity $ F\approx 0.858 $ for the high loss regime. Related studies~\cite{Hermans2023-kf} aim to generate states at a target fidelity close to 1. In a similar vein, we may optimize $ \gamma $ to obtain $ F\geq F_{\mathrm{target}} $ in lieu of choosing a $ \gamma $ that maximizes $ \mathcal{R}(\rho_{AB}) $. Fig.~\ref{fig:rate_targFid} compares the optimal $ \mathcal{R}(\rho_{AB}) $ (blue) with the cases where we set a target fidelity of $ 1-10^{-2}$ (orange), $ 1-10^{-3} $ (yellow), and $ 1-10^{-4} $ (purple). All the rate curves are parallel i.e. they maintain the $ \mathcal{O}(\sqrt{\eta}) $ scaling for $ \eta\ll1 $. However, the actual rates are lower than the optimal rate curve. This also corresponds to a longer range (i.e.\ higher overall loss) at which the single rail swap starts outperforming the dual rail swap (note the cross over between the solid and red dashed lines). 
Readers interested in detailed analysis of the $ \mathcal{R}(\rho_{AB}) $, $ P_{\mathrm{succ.}} $ and $ I(\rho_{AB}) $ of the states over the whole parameter space may refer to the detailed state analyses included in Appendices~\ref{app:detailed_state_deriv} -- \ref{app:general_state} of the manuscript.
	
	\section{Improvements using Entanglement Distillation}
	\label{sec:distillation}
\begin{figure}[htbp]
	\centering
	\includegraphics[width=0.8\linewidth]{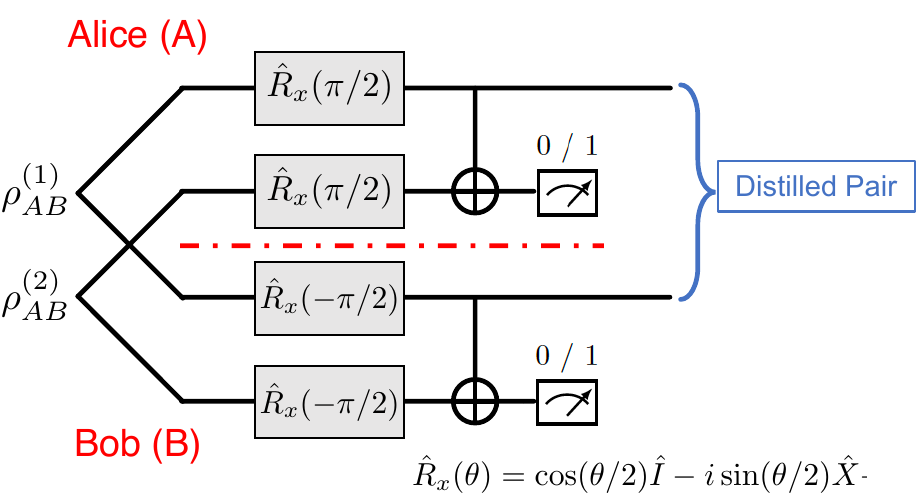}
	\caption{Distillation protocol intrdouced in~\cite{Deutsch1996-sb} using local $ \hat{R}_x(\pm\pi/2) $ gates, bilateral $ \mathrm{CNOT} $ gates, and Pauli-$ Z $ measurements. A `success' is declared when both measurement outcomes are `0' or `1', followed by reconciliation (using classical communication) between the two parties.  }
	\label{fig:distill_ckt}
\end{figure}

Hardware non-idealities limit the utility of entanglement generation using photonic BSMs as discussed in Sec.~\ref{sec:tradeoffs}. Entanglement distillation is one way to improve the utility, at the expense of additional quantum state processing at the end user ($A$ amd $B$) sites. As an example, we consider the distillation protocol introduced in~\cite{Deutsch1996-sb} and demonstrate its ability to improve state quality. { We omit any decoherence, dephasing, dissipation or erasure (between subsequent gate operations and for the duration of classical communication) in our quantum memories, since these effects are qubit-implementation specific and beyond the scope of our study. }Fig.~\ref{fig:distill_ckt}, highlights the the distillation process --- Alice (Bob) generate a pair of entangled states and perform $ \pi/2 (-\pi/2)$  rotations along the qubit $ x $-axis (represented by $ {\hat{R}_x}(\pm\pi/2) $) on both of their qubits. This is followed by bilateral $ \mathrm{CNOT} $ gates and Pauli-$ {Z} $  basis measurements on the target qubits. The distillation process is successful if both measurement outcomes concur.

\begin{figure*}[bpht!]
    \centering
    \includegraphics[width=0.85\linewidth]{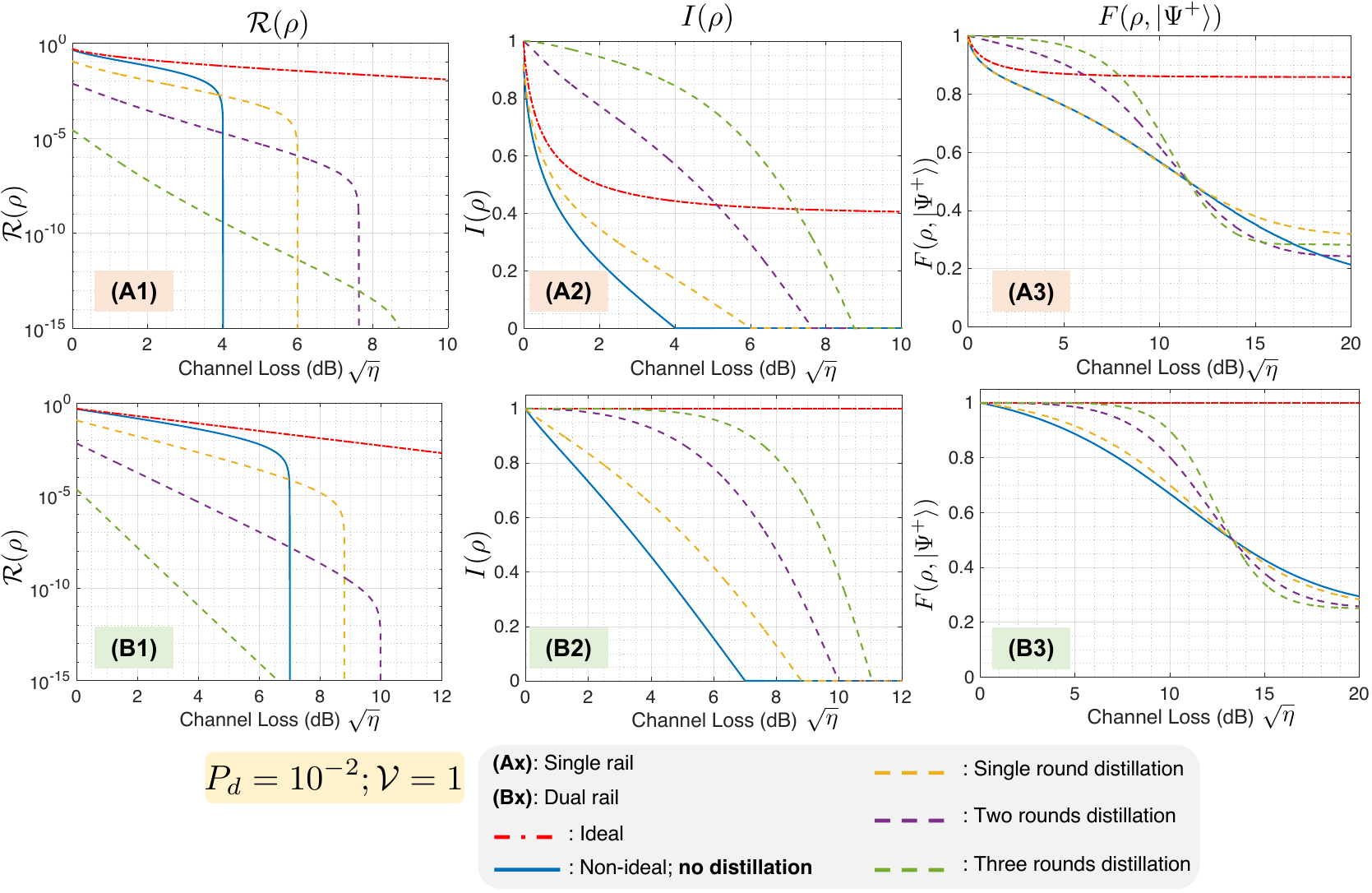}
	\caption{State quality evaluation after rounds of distillation heralded swaps with $ P_d=10^{-2} $ and $ \mathcal{V}=1 $. We plot the (A1,B1) hashing bound rate $ \mathcal{R}(\rho)$, (A2,B2) hashing bound per copy $ I(\rho) $ and, (A3,B3) fidelity of the distilled states $ F(\rho,\ket{\Psi^+}) $ (dashed lines) and compare it with the pre-distillation (blue solid) and ideally heralded (red dot-dashed) states for the single (top panes; A) and dual rail (bottom panes; B) encodings for entanglement swaps.}
	\label{fig:distill_plots}
\end{figure*}

Fig.~\ref{fig:distill_plots} summarizes the improvements in the heralded state quality under multiple rounds of distillation. We consider swaps with $ P_d=10^{-2} $ excess photons per mode and no photon distinguishability i.e.\ $\mathcal{V}=1$. 
The panes labeled (A1) and (B1) highlight the improvement in the maximum range of the un-distilled state, when we employ multiple rounds of distillation. This comes at the cost of a poorer rate scaling for each subsequent round of distillation.  Given the rate scaling of $ \mathcal{O}(\sqrt{\eta}) $ and $ \mathcal{O}({\eta}) $ for the single and dual rail swapping protocols respectively, with $ k $-rounds on distillation the rate scalings drop to $ \mathcal{O}(\eta^{2^{k-1}}) $ and $ \mathcal{O}(\eta^{2^{k}}) $ respectively. 

\begin{figure}[ht!]
	\centering
	\includegraphics[width=0.9\linewidth]{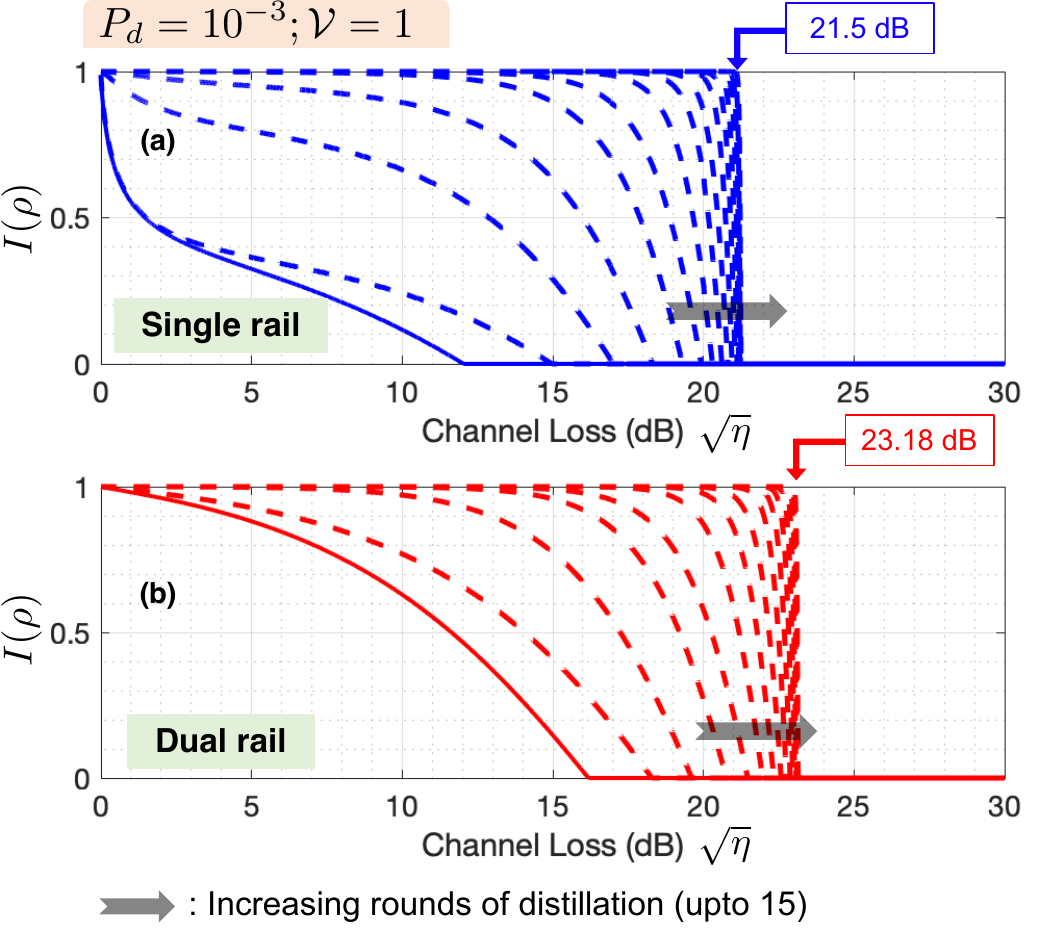}
	\caption{Improvement in the state $ I(\rho) $ from the heralded original state (solid lines) under multiple rounds of distillation (dashed lines; upto 15 rounds of distillation) for the (a) single rail and (b) dual rail case. The limiting value of maximum range $ \eta_{\lim} $ is labeled. Black arrow marks the direction for multiple rounds of distillation. We consider considering $ P_d=10^{-3}; \mathcal{V}=1; \varepsilon=0$ for both encoding choices.  }
	\label{fig:distill_limit}
\end{figure}

Given the rapid deterioration in the rate scaling, we consider the state's $ I(\rho) $ as shown in the panes labeled (A2) and (B2). This supports the conclusions drawn from the distillable entanglement rate discussion above. We note that under continued rounds of distillation, the max range is limited to a finite value of loss, which we label $ \eta_{\lim} $. Fig.~\ref{fig:distill_limit} highlights the limiting values for the single and dual rail cases considering $ P_d=10^{-3}; \mathcal{V}=1$.

It is important to note here that all the analysis in this article has been focussed on \emph{achievable} distillable entanglement. In the regime where $ I(\rho)\rightarrow 0 $, we cannot conclusively claim that about the actual entanglement content in the heralded and/or distilled state is indeed zero. A complete characterization could be performed by calculating an upper bound to the heralded state's distillable entanglement. However these quantities are hard to calculate for our generalized analytic state formulations; we leave this problem open for future studies. 

\begin{figure}[h!]
	\centering
	\includegraphics[width=0.9\linewidth]{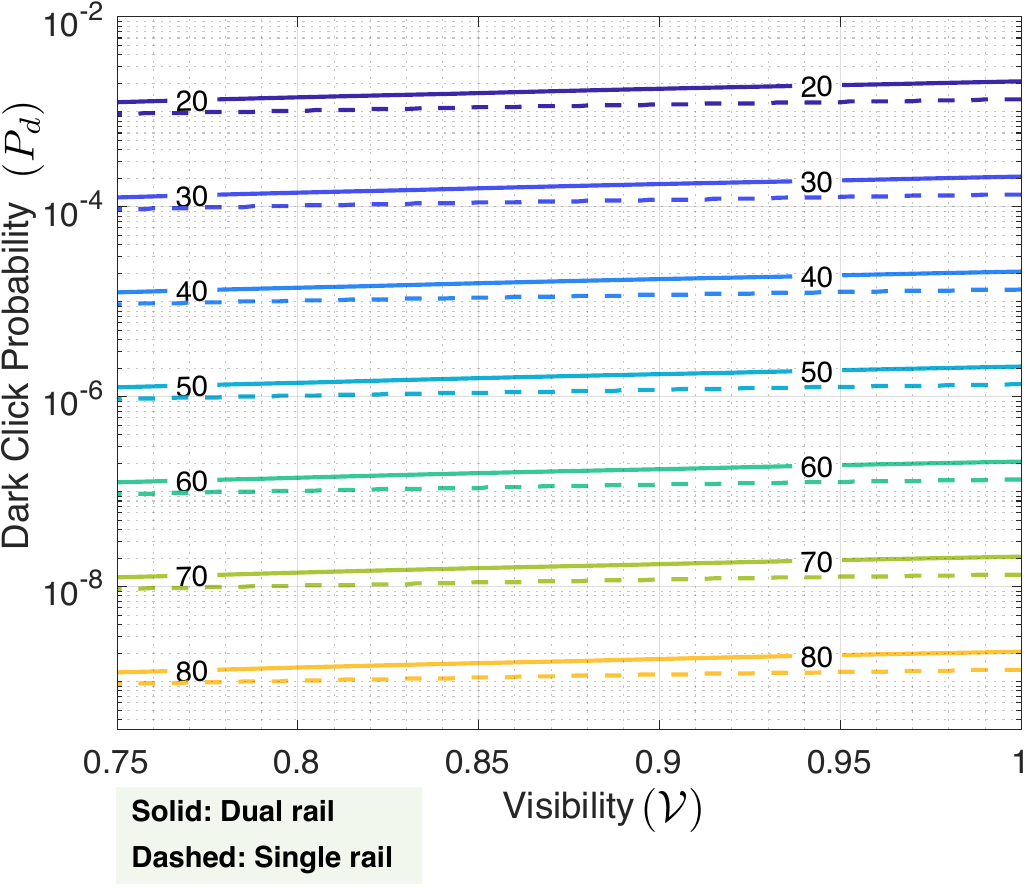}
	\caption{Contour plot for $ \eta_{\lim} $ (labels in dB) for varying values of excess noise $(P_d)  $ and visibility $(\mathcal{V}) 
 $ of the dual (solid) and single (dashed) rail swapped states.  }
	\label{fig:distill_contour}
\end{figure}

In panes (A3) and (B3), we analyze the fidelity of the distilled states. As expected the fidelity improves under distillation but only upto a specific value of channel loss where $ F(\rho_{AB},\ket{\Psi^+})=0.5  $. One may solve $ F(\rho_{AB},\ket{\Psi^+})=0.5  $ for $ \eta $ to evaluate the limiting range $ \eta_{\lim} $. The reson behind this is straightforward-- at  $ F(\rho_{AB},\ket{\Psi^+})=0.5 $ the initial joint state is the two-qubit maximally mixed state $ \rho_{AB}=\mathbb{I}/4 $. This state has no distillable entanglement i.e.\ $ E_D(\mathbb{I}/4)=0 $. Consequently, starting with this state, no distillation protocol will improve the state quality. For $ \eta<\eta_{\lim} $, the heralded state  $ F(\rho_{AB},\ket{\Psi^+})>0.5  $, meaning the state quality may be improved by distillation. Hence $ I(\rho) $ improves asymptotically (with additional rounds of distillation) to 1 for $ \eta<\eta_{\lim} $. We plot contours of $\eta_{\lim}$ (labelled in dB)  for various values of excess noise and state visibility in Fig.~\ref{fig:distill_contour}. We refer the reader to Appendix~\ref{app:distill} for more details. Additional analysis of more system specific distillation schemes (for e.g., as proposed in Ref.~\cite{Rozpedek2018-ll}) are potential topics for future studies.

	\section{Results and Conclusions}
	\label{sec:conclusion}

The general study of the `swap in the middle' entanglement distribution architecture carried out in this paper is relevant for various physical quantum memories, including color-center, trapped ion, neutral-atom and superconducting qubits. Our calculations highlight the key tradeoffs \emph{vis.\ a vis.\ }photonic-qubit encoding that need to be accounted for in such entanglement distribution among two memories connected by an optical channel, mediated by a heralded photonic swap. The evaluation of the distillable entanglement rate as compared to entangled-state distribution rate (at a given target fidelity) serves as an useful metric for the fair comparison of the two photonic-qubit encoding choices considered here. There are some pros and cons between the two encodings. Dual rail wins in the low loss regime, but has an inferior high-loss rate scaling. However, dual rail is unaffected by carrier-phase mismatch, and is less affected (than single-rail) by excess noise. Imperfect visibility in the swap appears to adversely effect both encodings similarly. Additionally, we find that using fidelity as the sole distributed state quality metric can lead to misleading and sub-optimal conclusions. Readers should note that, our claims do not preclude other encoding choices, i.e. there may be higher dimensional encodings that close the `gap' between the swap performance plots we show for the two encodings, and the repeaterless bound, even for the midpoint swap architecture, especially in the low-loss regime, which is relevant for short-range quantum links. We demonstrate the varying effect posed by the major sources {  affecting fidelity} in the final entangled state between the memory qubits. We hope that the results of our study will serve as illuminating guidelines for experimental demonstrations, and further research into advanced entanglement distillation protocols pursuant to the actual form of the heralded entangled state, as well as future extensions of our work to more complex quantum network topologies. 
	
	\section{Acknowledgments}
	We thank Kevin C. Chen (MIT), Eneet Kaur (Univ. of Arizona), Filip Rozpedek (Univ.\ of Chicago), Johannes Borregaard (TU Delft) and Babak N. Saif (GSFC, NASA) for fruitful discussions and helpful comments about the manuscript. The authors acknowledge the Mega Qubit Router (MQR) project funded under federal support via a subcontract from the University of Arizona Applied Research Corporation (UA-ARC). Additionally, the authors  acknowledge National Science Foundation (NSF) Engineering Research Center for Quantum Networks (CQN), awarded under cooperative agreement number 1941583, for supporting this research.

	 S.G. has outside interests in SensorQ Technologies Incorporated and Guha, LLC. D.E. holds shares in Quantum Network Technologies (QNT), a performer on the MQR award. These interests have been disclosed to UA and MIT respectively, and reviewed in accordance with their conflict of interest policies, with any conflicts of interest to be managed accordingly.
  
        \onecolumngrid
	
	\renewcommand\thefigure{\thesection\arabic{figure}}  
	\counterwithin{figure}{section}  
	\appendix 
	\section{Parameter Index}
Throughout this section we shall use the following common symbols. Wherever the subscript $ k $ is used, $ k=A $ and $ k=B $ refers to Alice's and Bob's parameters respectively.  We assume that all similar hardware components are identical i.e. all detectors are alike, all beamsplitters have the same configuration etc.

\begin{table}[h!]
	\begin{tabular}{>{\centering}p{4cm}>{}p{10cm}}
		\toprule
		\textbf{Symbol} & \textbf{Definition}\\\midrule
		${\eta_k}$ & Individual channel transmissivity from source $ k $ to Entanglement Swapping Station.  \\ 
		\midrule
		${\gamma_k}$ & Qubit initialization parameter for source $ k $ (only for single rail case). \\ 
		\midrule
		${\eta_d}$ & Detection efficiency of photon detectors.  \\ 
		\midrule
		${P_d}$ & Total excess noise (photons per qubit slot) in photon detectors.  \\ 
		\midrule
		${\mathcal{V}=|\mathcal{V}|\exp(i \arg \mathcal{V})}$ & Mode matching parameter for individual interacting photonic pulses with $ |\mathcal{V}| $ evaluates mode overlap and $ \arg\mathcal{V} $ evaluates the carrier phase mismatch. This parameter is squared for the dual rail case.  \\ 
		\midrule
		$F(\rho,\ket{\Psi^\pm})$ & Fidelity of density operator $ \rho $ with target Bell states $ \ket{\Psi^\pm} $  \\ 
		\midrule
		$P_{\mathrm{succ.}}$ & Entanglement swapping rate (at a given fidelity target whenever applicable)  \\ 
		\midrule
		$I(\rho_{kl} )$ & Distillable entanglement of the bi-partite state  $ \rho_{kl} $  \\ 
		\midrule
		$\mathcal{R}(\rho_{kl})$ & Lower bound to ultimate entanglement swapping rate to generate state : $ \mathcal{R}(\rho)=I (\rho)\times P_{\mathrm{succ.}} $  \\ 
		\bottomrule
	\end{tabular}
	\caption{\label{tab:parameters} Symbols with their corresponding definitions.}
\end{table}
	\section{Detailed Derivation of Entangled QM State}
\label{app:detailed_state_deriv}
\subsection{Encoding Choices}
The \emph{single rail} encoding, is defined by the presence or absence of a photon in a single bosonic mode; the logical qubit states for the mode labeled by $ k $ are defined by 
\begin{align}
	\ket{\bar{0}} \equiv \ket{0}_k ;\ket{\bar{1}}\equiv \ket{1}_k.  
\end{align}

In the \emph{dual rail} encoding, the logical qubit states are represented by the presence of a single photon in one of two orthogonal bosonic modes labeled by $ k1 $ and $ k2 $ (which may be spatial, temporal, spectral modes or any combination thereof), as
\begin{align}
	\begin{split}
		\ket{\bar{0}}\equiv\ket{1}_{k1}\ket{0}_{k2} =\ket{1,0}_{k};\\
		\ket{\bar{1}}\equiv\ket{0}_{k1}\ket{1}_{k2}=\ket{0,1}_k .
	\end{split}
\end{align}

\subsection{Single Rail }
\label{app:subsec_single_deriv} 
\textit{Initial State:} We start with the general initial state for a single rail photonic qubit (subscript $k$) entangled to a quantum memory (the kets with the bold font) held by the party denoted by the subscript $S=A,B$,  
\begin{align}
	\ket{\psi(\gamma)}_{S}=\sqrt{\gamma}\memk{1}_S\ket{0}_{S,k} +e^{i\theta_S} \sqrt{1-\gamma}\memk{0}_{S}\ket{1}_{S,k}.
\end{align}

\textit{State After Loss:}
The general state of the QM and single rail encoded photonic qubit after undergoing bosonic pure loss of magnitude $\eta$ on the photonic mode is given by,
\begin{align}
    \ket{\psi(\gamma,\eta,\theta_S)}_S=\sqrt{\gamma}\memk{1}_{S}\ket{0}_{S,k}\otimes \ket{0}_E +e^{i\theta_S}\sqrt{1-\gamma}\sqrt{\eta}\memk{0}_{S}\ket{1}_{S,k}\otimes \ket{0}_E+
	e^{i\theta_S}\sqrt{1-\gamma}\sqrt{1-\eta}\memk{0}_{S}\ket{0}_{S,k} \otimes \ket{1}_E.
	\label{eq: ent_st_sing1}
\end{align}
The subscript $E$ denotes the environment modes that we shall trace out subsequently.

\textit{State after beamsplitter:} For the beamsplitter interaction, we have to consider two copies of entangled states in Eq.~\eqref{eq: ent_st_sing1}, with $S=A,B$ respectively. After the standard balanced beamsplitter interaction, the output state has three main components -- (1) where no photons will be detected by ideal detectors, (2) only one photon will be detected, and (3) more than one photon is detected. These components are respectively given as ---

\begin{subequations}
\noindent No photons detected: 
 \begin{align}
 		\begin{split}
 			\ket{\varphi}_{s,0}=U_{BS} (\ket{\psi}_A\otimes \ket{\psi}_B)^{\mathrm{0- photon}}=&(\sqrt{\gamma_A}\memk{1}_{A}\ket{0}_{A,k}\otimes \ket{0}_E +
 		e^{i\theta_A}\sqrt{1-\gamma_A}\sqrt{1-\eta_A}\memk{0}_{A}\ket{0}_{A,k} \otimes \ket{1}_E)\\
 		& \otimes (\sqrt{\gamma_B}\memk{1}_{B}\ket{0}_{B,k}\otimes \ket{0}_E +
 		e^{i\theta_B}\sqrt{1-\gamma_B}\sqrt{1-\eta_B}\memk{0}_{B}\ket{0}_{B,k} \otimes \ket{1}_E)
 	\end{split}
 \end{align}
Single photon detected: 
\begin{align}
	\begin{split}
	\ket{\varphi}_{s,1}=&	U_{BS} (\ket{\psi}_A\otimes \ket{\psi}_B)^{\mathrm{1- photon}}\\
        = &\frac{1}{\sqrt{2}}e^{i\theta_A}{\sqrt{1-\gamma_A}\sqrt{\eta_A}}\times{\sqrt{\gamma_B}}\memk{0}_A\memk{1}_B\Bigl[ \ket{1}_{A,k} \ket{0}_{B,k}+ \ket{0}_{A,k} \ket{1}_{B,k}\Bigr] \otimes \ket{0,0}_E \\
		&+\frac{1}{\sqrt{2}} e^{i\theta_B} {\sqrt{1-\gamma_B}\sqrt{\eta_B}}\times{\sqrt{\gamma_A}}\memk{1}_A\memk{0}_B\Bigl[ -\ket{1}_{A,k} \ket{0}_{B,k}+ \ket{0}_{A,k} \ket{1}_{B,k}\Bigr] \otimes \ket{0,0}_E \\
		&+\frac{1}{\sqrt{2}}e^{i(\theta_A+\theta_B)}{\sqrt{1-\gamma_A}\sqrt{\eta_A}}{\sqrt{1-\gamma_B}\sqrt{1-\eta_B}} \memk{0}_A\memk{0}_B \Bigl[ \ket{1}_{A,k} \ket{0}_{B,k}+ \ket{0}_{A,k} \ket{1}_{B,k}\Bigr]\otimes \ket{0,1}_E \\
		&+\frac{1}{\sqrt{2}}e^{i(\theta_A+\theta_B)}{\sqrt{1-\gamma_A}\sqrt{1-\eta_A}}{\sqrt{1-\gamma_B}\sqrt{\eta_B}}\memk{0}_A\memk{0}_B \Bigl[ -\ket{1}_{A,k} \ket{0}_{B,k}+ \ket{0}_{A,k} \ket{1}_{B,k}\Bigr]\otimes \ket{1,0}_E \\
	\end{split}
\end{align}
More than one photon detected:
\begin{align}
	\ket{\varphi}_{s,2}&= U_{BS} (\ket{\psi}_A\otimes \ket{\psi}_B)^{\mathrm{2- photon}} \nonumber\\
 &= e^{i(\theta_A+\theta_B)}\sqrt{(1-\gamma_A)(1-\gamma_B)}\times \sqrt{\eta_A\eta_B}\memk{0}_A\memk{0}_B[\ket{0}_{A,k}\ket{2}_{B,k}- \ket{0}_{A,k}\ket{2}_{B,k}]\otimes\ket{0,0}_E
\end{align}
\end{subequations}

The final state is obtained by projecting the photonic modes $A,k \text{ and } B,k$ to the states $\ket{0,1}$ or $\ket{1,0}$. We then trace out the environment modes to obtain the final spin-spin state,
\begin{align}
	{\rho}_{\mathrm{single}}= (1-P_d)^2\Tr_E(\outprod{\varphi}_{s,1})+(1-P_d)P_d\Tr_E(\outprod{\varphi}_{s,0})
\end{align}
where,
\begin{align}
    \begin{split}
        \Tr_E(\outprod{\varphi}_{s,1})= &\frac{1}{2} \left( e^{i\theta_A}{\sqrt{1-\gamma_A}\sqrt{\eta_A}}\times{\sqrt{\gamma_B}}\memk{0}_A\memk{1}_B+ (-1)^{m_1} e^{i\theta_B} {\sqrt{1-\gamma_B}\sqrt{\eta_B}}\times{\sqrt{\gamma_A}}\memk{1}_A\memk{0}_B \right) \\
    &\otimes \left( e^{-i\theta_A}{\sqrt{1-\gamma_A}\sqrt{\eta_A}}\times{\sqrt{\gamma_B}}\memb{0}_A\memb{1}_B+ (-1)^{m_1} e^{-i\theta_B} {\sqrt{1-\gamma_B}\sqrt{\eta_B}}\times{\sqrt{\gamma_A}}\memb{1}_A\memb{0}_B \right)\\
    &+ \frac{1}{2}\left[(1-\gamma_A)(1-\gamma_B)(\eta_A+\eta_B-2\eta_A\eta_B)\right] \memk{0}_A\memk{0}_B\memb{0}_A\memb{0}_B,
    \end{split}
    \label{eq:sing_oneph}
\end{align}
with $m_1=0$ for the $[0,1]$ click pattern and $m_1=1$ for the $[1,0]$ click pattern and,
\begin{align}
    \begin{split}
         \Tr_E(\outprod{\varphi}_{s,0})=& \; \gamma_A\gamma_B \memk{1}_A\memk{1}_B\memb{1}_A\memb{1}_B +(1-\gamma_A)(1-\gamma_B)(1-\eta_A)(1-\eta_B)\memk{0}_A\memk{0}_B\memb{0}_A\memb{0}_B\\
         &+\gamma_B(1-\gamma_A)(1-\eta_A)\memk{0}_A\memk{1}_B\memb{0}_A\memb{1}_B +\gamma_A(1-\gamma_B)(1-\eta_B)\memk{1}_A\memk{0}_B\memb{1}_A\memb{0}_B.
    \end{split}
\end{align}

\subsection{Dual Rail }
\label{app:subsec_dual_deriv}
\textit{Initial State:} We start with the general initial state for a dual rail photonic qubit (subscript $k$ and the kets with overbars) entangled to a quantum memory (the kets with the bold font) held by the party denoted by the subscript $S=A,B$,  
\begin{align}
	\ket{\psi}_{S}=e^{i\theta_S}\left[\frac{1}{\sqrt{2}}\memk{1}_S\ket{\bar{0}}_{S,k} +\frac{1}{\sqrt{2}}\memk{0}_{S}\ket{\bar{1}}_{S,k}\right]=e^{i\theta_S}\left[\frac{1}{\sqrt{2}}\memk{1}_S\ket{1,0}_{S,k} +\frac{1}{\sqrt{2}}\memk{0}_{S}\ket{0,1}_{S,k}\right]   ,
\end{align}

\textit{State after loss:} The general state of the QM and dual rail encoded photonic qubit after undergoing bosonic pure loss of magnitude $\eta$ (for each underlying mode) is given by,
\begin{align}
	\begin{split}
			\ket{\psi (\eta,\theta_S))}_{S}= &e^{i\theta_S}\frac{1}{\sqrt{2}} 	\sqrt{\eta}\memk{1}_S\ket{1,0}_{S,k}\ket{0,0}_E +e^{i\theta_S} \frac{1}{\sqrt{2}}\sqrt{\eta} \memk{0}_{S} \ket{0,1}_{S,k}\ket{0,0}_E  \\
		&+ e^{i\theta_S}\frac{1}{\sqrt{2}} 		\sqrt{1-\eta}\memk{1}_S\ket{0,0}_{S,k}\ket{1,0}_E +e^{i\theta_S} \frac{1}{\sqrt{2}}\sqrt{1-\eta} \memk{0}_{S} \ket{0,0}_{S,k}\ket{0,1}_E  ,
		\label{eq:ent_st_dual1}
	\end{split}
\end{align}
The subscript $E$ denotes the environment modes that we shall trace out subsequently.

\textit{State after beamsplitter:} For the beamsplitter interaction, we have to consider two copies of entangled states in Eq.~\eqref{eq:ent_st_dual1}. Considering two standard balanced beamsplitters, the output state has four main components (1) no photons will be detected by ideal detectors (2) at most one photon is detected by a single detector, (3) two detectors detect a single photon each, and (4) multi-photon detection events are observed. These components are respectively given as ---

\begin{subequations}
\noindent No photons are detected:
\begin{align}
\begin{split}
		\ket{\varphi}_{d,0}= & e^{i\theta_A}(\frac{1}{\sqrt{2}} \sqrt{1-\eta_A}\memk{1}_A\ket{0,0}_{A,k}\ket{1,0}_E + \frac{1}{\sqrt{2}}\sqrt{1-\eta_A} \memk{0}_{A} \ket{0,0}_{A,k}\ket{0,1}_E )\\
	&\otimes 
	e^{i\theta_B}(\frac{1}{\sqrt{2}} \sqrt{1-\eta_B}\memk{1}_B  \ket{0,0}_{B,k}\ket{1,0}_E + \frac{1}{\sqrt{2}}\sqrt{1-\eta_B} \memk{0}_{B} \ket{0,0}_{B,k}\ket{0,1}_E )
	\end{split}
\end{align}
At most one photon is detected:
\begin{align}
	\begin{split}
		\ket{\varphi}_{d,1}=
		\frac{e^{i(\theta_A+\theta_B)}}{2\sqrt{2}}\biggl[& \sqrt{\eta_A(1-\eta_B)}\memk{1}_A \memk{1}_B \bigl[\ket{1,0}_{A,k}  \ket{0,0}_{B,k}+ \ket{0,0}_{A,k}  \ket{1,0}_{B,k}\bigr] \ket{0,0,1,0}_E \\
        +& \sqrt{\eta_A(1-\eta_B)}\memk{1}_A \memk{0}_B \bigl[\ket{1,0}_{A,k}  \ket{0,0}_{B,k} + \ket{0,0}_{A,k}  \ket{1,0}_{B,k}\bigr] \ket{0,0,0,1}_E \\
        +& \sqrt{\eta_A(1-\eta_B)}\memk{0}_A \memk{1}_B \bigl[\ket{0,1}_{A,k}  \ket{0,0}_{B,k}+\ket{0,0}_{A,k}  \ket{0,1}_{B,k}\bigr] \ket{0,0,1,0}_E \\
        +& \sqrt{\eta_A(1-\eta_B)}\memk{0}_A \memk{0}_B \bigl[\ket{0,1}_{A,k}  \ket{0,0}_{B,k}+\ket{0,0}_{A,k}  \ket{0,1}_{B,k}\bigr] \ket{0,0,0,1}_E \\
		+& \sqrt{\eta_B(1-\eta_A)}\memk{1}_A \memk{1}_B \bigl[\ket{0,0}_{A,k}  \ket{1,0}_{B,k}-\ket{1,0}_{A,k}  \ket{0,0}_{B,k}\bigr] \ket{1,0,0,0}_E \\
         +& \sqrt{\eta_B(1-\eta_A)}\memk{0}_A \memk{1}_B \bigl[\ket{0,0}_{A,k}  \ket{1,0}_{B,k}-\ket{1,0}_{A,k}  \ket{0,0}_{B,k}\bigr] \ket{0,1,0,0}_E \\
         +& \sqrt{\eta_B(1-\eta_A)}\memk{1}_A \memk{0}_B \bigl[\ket{0,0}_{A,k}  \ket{1,0}_{B,k}-\ket{1,0}_{A,k}  \ket{0,0}_{B,k}\bigr] \ket{1,0,0,0}_E \\
		+& \sqrt{\eta_B(1-\eta_A)}\memk{0}_A \memk{0}_B \bigl[\ket{0,0}_{A,k}  \ket{1,0}_{B,k}-\ket{1,0}_{A,k}  \ket{0,0}_{B,k}\bigr] \ket{0,1,0,0}_E \biggr]
	\end{split}
\end{align}
Single photons are detected by a pair of detectors:
\begin{align}
	\begin{split}
		\ket{\varphi}_{d,2}=
		&\frac{e^{i(\theta_A+\theta_B)}}{4} \sqrt{\eta_A\eta_B} \biggl[\memk{1}_A \memk{0}_B \otimes \bigl[\ket{1,0}_{A,k}  \ket{0,1}_{B,k}-\ket{1,1}_{A,k}  \ket{0,0}_{B,k} + \ket{0,0}_{A,k}  \ket{1,1}_{B,k}-\ket{0,1}_{A,k}  \ket{1,0}_{B,k} \bigr]\\
		&+ \memk{0}_A \memk{1}_B \otimes \bigl[-\ket{1,0}_{A,k}  \ket{0,1}_{B,k}-\ket{1,1}_{A,k}  \ket{0,0}_{B,k} + \ket{0,0}_{A,k}  \ket{1,1}_{B,k}+\ket{0,1}_{A,k}  \ket{1,0}_{B,k} \bigr]  \biggr]\otimes \ket{0,0,0,0}_E\\
	\end{split}
\end{align}
Multiple photons are detected by the same detector. 
\begin{align}
\begin{split}
\ket{\varphi}_{d,3}=\frac{e^{i(\theta_A+\theta_B)}}{2} \sqrt{\eta_A\eta_B} \biggl[ &\memk{1}_A \memk{1}_B \bigl[\ket{2,0}_{A,k}  \ket{0,0}_{B,k}+\ket{0,0}_{A,k}  \ket{2,0}_{B,k} \bigr] \\
		&+\memk{0}_A \memk{0}_B \bigl[\ket{0,2}_{A,k}  \ket{0,0}_{B,k}+\ket{0,0}_{A,k}  \ket{0,2}_{B,k} \bigr] \biggr] \otimes \ket{0,0,0,0}_E 
\end{split}
\end{align}
\end{subequations}

The final state is obtained by projecting the photonic modes $A,k \text{ and } B,k$ to the states $\ket{0,0,1,1},\,\ket{0,1,1,0},\,\ket{1,0,0,1}$ or $\ket{1,1,0,0}$. We then trace out the environment modes to obtain the final spin-spin state,
\begin{align}
	{\rho}_{\mathrm{dual}}= (1-P_d)^4\Tr_E(\outprod{\varphi}_{d,2})+(1-P_d)^3 P_d\Tr_E(\outprod{\varphi}_{d,1})+(1-P_d)^2 P_d^2 \Tr_E(\outprod{\varphi}_{d,0})
\end{align}
where,
\begin{align}
    \Tr_E(\outprod{\varphi}_{d,2})= \frac{\eta_A\eta_B}{16}\left(\memk{0}_A \memk{1}_B+(-1)^{m_1}\memk{1}_A \memk{0}_B\right)\left(\memb{0}_A \memb{1}_B+(-1)^{m_1}\memb{1}_A \memb{0}_B\right)
    \label{eq:dual_rail_singph}
\end{align}
with $m_1=0$ for the click patterns $[1,1,0,0]$ and $[0,0,1,1]$, and $m_1=1$ for the click patterns $[1,0,0,1]$ and $[0,1,1,0]$, and 
{\small
\begin{align}
    \begin{split}
        \Tr_E(\outprod{\varphi}_{d,1})=\frac{1}{8}\times(\eta_A+\eta_B-2\eta_A\eta_B) \biggl(& \memk{1}_A\memk{1}_B\memb{1}_A\memb{1}_B +\memk{1}_A\memk{0}_B\memb{1}_A\memb{0}_B + \memk{0}_A\memk{1}_B\memb{0}_A\memb{1}_B + \memk{0}_A\memk{0}_B\memb{0}_A\memb{0}_B\biggr),
    \end{split}
\end{align}
}
{\small 
\begin{align}
    \Tr_E(\outprod{\varphi}_{d,0})=\frac{1}{4}(1-\eta_A)(1-\eta_B)\biggl(& \memk{1}_A\memk{1}_B\memb{1}_A\memb{1}_B +\memk{1}_A\memk{0}_B\memb{1}_A\memb{0}_B + \memk{0}_A\memk{1}_B\memb{0}_A\memb{1}_B + \memk{0}_A\memk{0}_B\memb{0}_A\memb{0}_B\biggr).
\end{align}
}

\subsection{Barrett-Kok Extension to the Single Rail Protocol}
Ref.~\cite{Barrett2004-sj} studies an extension to the single-rail photonic qubit assited swapping protocol by using a second emission from the memories to `correct' for the error term $\memk{0}_A\memk{0}_B\memb{0}_A\memb{0}_B$. Let us consider the case for $P_d=0$. Analyzing the state of the memories, we first apply a logical $\mathrm{NOT}$ operation on the terms of Eq.~\eqref{eq:sing_oneph}, and generate memory photon entanglement a second time. This yields the state,
\begin{align}
    \begin{split}
     \sigma_{1}   = &\frac{1}{2} e^{i(\theta_A+\theta_B)}\left( {\sqrt{1-\gamma_A}\sqrt{\eta_A}}\times{\sqrt{\gamma_B}}\memk{1}_A\memk{0}_B\ket{0}_{A,k}\ket{1}_{B,k}+ (-1)^{m_1} {\sqrt{1-\gamma_B}\sqrt{\eta_B}}\times{\sqrt{\gamma_A}}\memk{0}_A\memk{1}_B \ket{1}_{A,k}\ket{0}_{B,k} \right) \\
    &\otimes e^{-i(\theta_A+\theta_B)} \left( {\sqrt{1-\gamma_A}\sqrt{\eta_A}}\times{\sqrt{\gamma_B}}\memb{1}_A\memb{0}_B \bra{0}_{A,k}\bra{1}_{B,k}+ (-1)^{m_1}{\sqrt{1-\gamma_B}\sqrt{\eta_B}}\times{\sqrt{\gamma_A}}\memb{0}_A\memb{1}_B \bra{1}_{A,k}\bra{0}_{B,k} \right)\\
    &+ \frac{1}{2}\left[(1-\gamma_A)(1-\gamma_B)(\eta_A+\eta_B-2\eta_A\eta_B)\right] \memk{1}_A\memk{1}_B\memb{1}_A\memb{1}_B \otimes\ket{0}_{A,k}\ket{0}_{B,k} \bra{0}_{A,k}\bra{0}_{B,k},
    \end{split}
\end{align}
Applying loss to the emitted photons, tracing out the environment, and detecting the additional photons for a click pattern of $[0,1]$ or $[1,0]$ yields the dual rail state in Eq.~\eqref{eq:dual_rail_singph}.

	\section{General Description of Final Spin-Spin Entangled State }
\label{app:general_state}
The generalized state description for both the single rail and dual rail encoded photonic entanglement swap is covered in this section. We state the density matrices of the final quantum memory (QM) entangled state with the two qubit basis ordering of $ \left\{ \memk{1,1},\memk{1,0} ,\memk{0,1}, \memk{0,0}  \right\} $. For the single rail case, we start with the initial QM- photon state of  $\sqrt{\gamma}\memk{1}\ket{0}+\sqrt{1-\gamma}\memk{0}\ket{1}$, whereas for the dual rail swap, we consider the initial state of $(\memk{1}\ket{1,0}+\memk{0}\ket{0,1})/\sqrt{2}$. 

\subsection{Single rail photonic qubit based swap}
\label{app:subsec_single_desc}
For the single rail case, the swapping circuit comprises of a single balanced beam splitter and a pair of photon number resolving detectors. A `success' is defined as a $ [0,1] $ or $ [1,0] $ click pattern, and the final entangled state's density operator in the chosen basis  is then given as, 
\begin{align}
	{\rho}_{\mathrm{single}}=\frac{(1-P_d)^2 }{\mathbf{N}_s} \begin{pmatrix}
		0 &  0 & 0 & 0 \\
		0 &  c_1^{(0)} & c_3^{(0)}  & 0 \\
		0 &  c_3^{(0)*}  &  c_2^{(0)} &0 \\
		0 & 0 & 0 & c_4^{(0)} \\
	\end{pmatrix} +\frac{P_d(1-P_d) }{\mathbf{N}_s} \begin{pmatrix}
		c_5^{(1)}&  0 & 0 & 0 \\
		0 &  c_1^{(1)} & 0& 0 \\
		0 & 0  &  c_2^{(1)} &0 \\
		0 & 0 & 0 & c_4^{(1)} \\
	\end{pmatrix} 
 \label{eq:dens_single_rail}
\end{align}
with $ \mathbf{N}_s=(1-P_d)^2 \left[ c_1^{(0)}+c_2^{(0)}+c_4^{(0)}\right]+P_d (1-P_d) \left[ c_1^{(1)}+c_2^{(1)}+c_4^{(1)}+c_5^{(1)} \right] $,  where,
\begin{align}
	\begin{split}
		c_1^{(0)}&=\frac{1}{2} \,\gamma_A \,(1 - \gamma_B) \,\eta_B \eta_d, \quad
		c_2^{(0)}=\frac{1}{2} \, (1 - \gamma_A) \, \gamma_B \, \eta_A \eta_d\\
		c_3^{(0)}&=\frac{1}{2} \, (-1)^m \eta_d \sqrt{\gamma_A(1 - \gamma_A)} \sqrt{\gamma_B(
			1 - \gamma_B)}  \sqrt{\eta_A \eta_B } \times \mathcal{V}\\
		c_4^{(0)}&=  \frac{1}{2} \eta_d  (1-\gamma_A)(1-\gamma_B)\left(\eta_A+\eta_B-2\eta_A\eta_B\eta_d\right)
	\end{split}
 \label{eq:sing_rail_st}
\end{align}
where $ m =\{0,1\}$ is a single parity bit determined by the click pattern - $m=0 \text{ for } [0,1];  m=1 \text{ for } [1,0]$. The visibility is calculated by  $\mathcal{V}=|\mathcal{V}|\exp(i(\theta_A-\theta_B))$, where the $|\mathcal{V}|$ is the mode overlap and $\theta_A,\theta_B$ are the individual carrier phases acquired by the photons from each source. The additional terms are expressible as,
\begin{align}
	\begin{split}
		c_1^{(1)}&= \gamma_A (1 - \gamma_B) (1 - \eta_B \eta_d); \quad
		c_2^{(1)}= (1 - \gamma_A) \gamma_B (1 - \eta_A \eta_d);\\
		c_4^{(1)}&=   (1 - \gamma_A) (1 - \gamma_B) (1 - \eta_A \eta_d) (1 - \eta_B \eta_d);\\
		c_5^{(1)}&=\gamma_A\gamma_B.
	\end{split}
\end{align}
We may express the visibility in a slightly different format as $\mathcal{V}=|\mathcal{V}|\exp(i\theta')$, where $\theta'=|\theta_A-\theta_B|$. Since both $\theta_A$ and $\theta_B$ are normal random variable with variance $\varepsilon$, i.e.\ $\theta_A,\theta_B \sim \mathcal{N}(0,\varepsilon) $, their difference is also anormal random variable of variance $2\varepsilon$, i.e.\ $\theta' \sim \mathcal{N}(0,2\varepsilon) $. 

To get an accurate description of the average state under the effect of random phase mismatch, we have to consider the ensemble averaged state. We parameterize each copy of $\rho_{\mathrm{single}}$ from Eq.~\eqref{eq:sing_rail_st} as $\rho_{\mathrm{single}}(\theta')$. Under this parameterization, we may calculate the ensemble averaged state as
\begin{align}
\rho_\mathrm{single}^{\mathrm{avg.}}=\int_{-\infty}^{\infty}\rho_\mathrm{single}(\theta)\cdot P(\theta) d\theta
\label{eq:ensemble_avg}
\end{align}
Since the only density operator term with $\theta'$ dependence is $c_3^{(0)}$, the ensemble averaged state $\rho_\mathrm{single}^{\mathrm{avg.}}$ has the same form as $\rho_\mathrm{single}$ with the modified density matrix entry,
\begin{align}
    c_3^{(0)}&=\frac{1}{2} \, (-1)^m \eta_d \sqrt{\gamma_A(1 - \gamma_A)} \sqrt{\gamma_B(
			1 - \gamma_B)}  \sqrt{\eta_A \eta_B } \times |\mathcal{V}| \exp(-\varepsilon)
\end{align}
where we have applied 
\begin{align}
    \int_{-\infty}^{\infty} e^{i\theta} \cdot \frac{\exp(-\theta^2/2\sigma^2) }{\sigma\sqrt{2\pi}} d\theta =\exp(-\sigma^2/2).
\end{align}
to resolve the integral in Eq.~\eqref{eq:ensemble_avg}.

\begin{figure}[ht]
	\centering
	\includegraphics[width=0.6\linewidth]{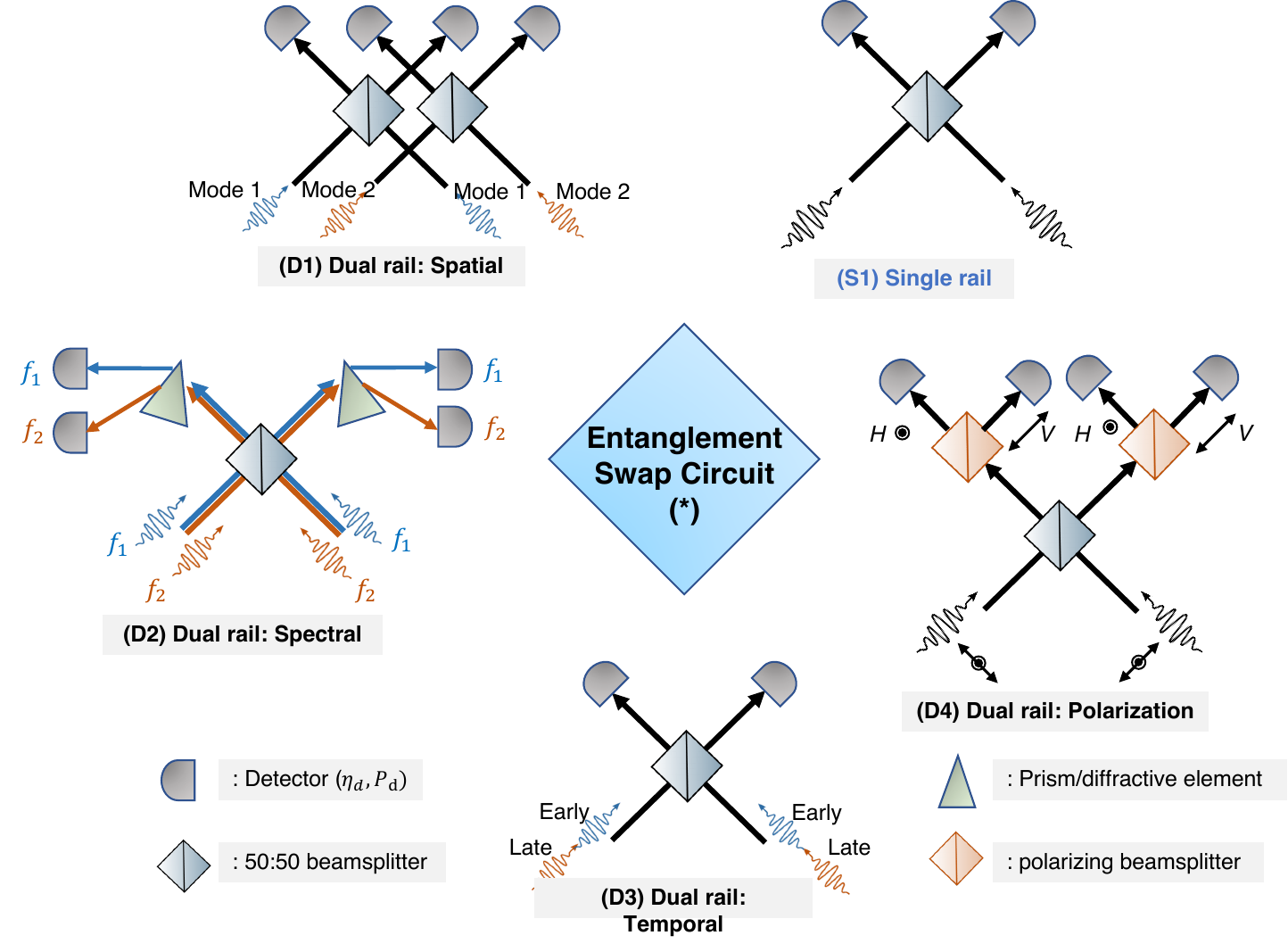}
	\caption{Various physical implementations of linear optical entanglement swaps based on the encoding choice. A single balanced beamsplitter and a pair of detectors is sufficient for the single rail swap (S1). For a dual rail spatial encoding requires two beamsplitters and four detectors (D1). A single beam splitter followed by a diffractive element is sufficient for spectrally encoded qubits (D2), whereas for polarization encodings a pair of polarizing beamsplitters are necessary (D4) each employing four detectors. For temporally encoded qubits, a single beamsplitter and a pair of detectors are sufficient (D3). }
	\label{fig:swaps}
\end{figure}

\subsection{Dual rail photonic qubit based swap}
\label{app:subsec_dual_desc}
For the dual rail case, the swapping circuit is dependent on the choice of the two orthogonal modes - we dpecit the hardware required for each encoding swap in Fig.~\ref{fig:swaps}. For spatial encoding, two balanced beam splitters and four detectors are required. For time-bin encoding, the swap comprises of a single balanced beamsplitter and two detectors (with 4 detection slots). For polarization encoding, we require a single balanced  beamsplitter, followed by two polarizing beamsplitters and four detectors. For spectral encoding the polarizing beam splitter is replaced by a diffractive or spectral de-multiplexing element.  There are four click patterns (of 8) that herald a successful swap.  The final entangled state's density operator in the chosen basis  is then given as, 

\begin{align}
	{\rho}_{\mathrm{dual}}= \frac{(1-P_d)^4 }{\mathbf{N}_d}\begin{pmatrix}
		0 & 0 & 0& 0\\
		0 & d_1^{(0)} & d_3^{(0)}& 0\\
		0 & d_3^{(0)*} & d_2^{(0)} &0\\
		0 & 0 & 0 & 0
	\end{pmatrix} + \frac{P_d(1-P_d)^2 }{\mathbf{N}_d}\begin{pmatrix}
		d_1^{(1)} & 0 & 0& 0\\
		0 & d_1^{(1)} & 0 & 0\\
		0 & 0 & d_1^{(1)} &0\\
		0 & 0 & 0 & d_1^{(1)}
	\end{pmatrix}
 \label{eq:dens_dual_rail}
\end{align}
with $ \mathbf{N}_d=2(1-P_d)^4 \times  d_1^{(0)} + 4P_d(1-P_d)^2 d_1^{(1)}$ where, 
\begin{align}
	d_{1}^{(0)}=d_{2}^{(0)}= \frac{1}{4} \eta_A \eta_B\eta_d^2; \quad
	d_{3}^{(0)}=\frac{1}{4} (-1)^m \eta_A \eta_B\eta_d^2\times |\mathcal{V}|^2,
\end{align}
where $ m =\{0,1\}$ is a single parity bit determined by the click pattern $(m=0 \text{ for } [0,1,1,0] \text{ or }[1,0,0,1] ;  m=1 \text{ for } [1,1,0,0] \text{ or } [0,0,1,1])$, and,
\begin{align}
	d_{1}^{(1)}=\frac{1}{2} (1-P_d)  \eta_d \left(\eta_A+\eta_B -2\eta_A \eta_B\eta_d\right) + P_d  (1-\eta_A \eta_d) (1-\eta_B \eta_d). 
\end{align}

\subsection{State Representation in the Bell basis}
\label{app:subsec_bell_desc}
An alternate representation of the density matrices for the entangled QMs is possible in the Bell state basis, i.e., $\ket{\Phi^\pm}=(\memk{0,0}\pm \memk{1,1})/\sqrt{2}$ and $\ket{\boldsymbol{\Psi}^\pm}=(\memk{0,1}\pm \memk{1,0})/\sqrt{2}$. Our basis ordering for the matrix representation is $\{\memk{\Psi^+}, \memk{\Psi^-}, \memk{\Phi^+}, \memk{\Phi^-}\}$. Consequently, the single rail state in Eq.~\eqref{eq:dens_single_rail} can be represented as,
\begin{align}
	{\rho}_{\mathrm{single}}=&\frac{(1-P_d)^2 }{2\mathbf{N}_s} \begin{pmatrix}
		c_1^{(0)}+c_2^{(0)}+c_3^{(0)}+c_3^{(0)*} &  -c_1^{(0)}+c_2^{(0)}+c_3^{(0)}-c_3^{(0)*} & 0 & 0 \\
		-c_1^{(0)}+c_2^{(0)}-c_3^{(0)}+c_3^{(0)*} &  c_1^{(0)}+c_2^{(0)}-c_3^{(0)}-c_3^{(0)*}& 0 & 0 \\
		0 &  0  &  c_4^{(0)} & c_4^{(0)}\\
		0 & 0 & c_4^{(0)} & c_4^{(0)} \\
	\end{pmatrix} \nonumber \\
& +\frac{P_d(1-P_d) }{2\mathbf{N}_s} \begin{pmatrix}
		c_1^{(1)} + c_2^{(1)} &  c_2^{(1)}-c_1^{(1)} & 0 & 0 \\
		c_2^{(1)}-c_1^{(1)} &  c_1^{(1)} + c_2^{(1)} & 0& 0 \\
		0 & 0  &  c_4^{(1)} + c_5^{(1)} & c_4^{(1)} - c_5^{(1)} \\
		0 & 0 & c_4^{(1)} - c_5^{(1)} & c_4^{(1)} + c_5^{(1)} \\
	\end{pmatrix} ,
  \label{eq:dens_single_railv2}
\end{align}
 and the dual rail state in Eq.~\eqref{eq:dens_dual_rail} is represented as,
 \begin{align}
	{\rho}_{\mathrm{dual}}=&\frac{(1-P_d)^2 }{2\mathbf{N}_d} \begin{pmatrix}
		d_1^{(0)}+d_2^{(0)}+d_3^{(0)}+d_3^{(0)*} &  -d_1^{(0)}+d_2^{(0)}+d_3^{(0)}-d_3^{(0)*} & 0 & 0 \\
		-d_1^{(0)}+d_2^{(0)}-d_3^{(0)}+d_3^{(0)*} &  d_1^{(0)}+d_2^{(0)}-d_3^{(0)}-d_3^{(0)*}& 0 & 0 \\
		0 &  0  &  0 & 0\\
		0 & 0 & 0 & 0 \\
	\end{pmatrix} \nonumber \\
 &+ \frac{P_d(1-P_d)^2 }{\mathbf{N}_d}\begin{pmatrix}
		d_1^{(1)} & 0 & 0& 0\\
		0 & d_1^{(1)} & 0 & 0\\
		0 & 0 & d_1^{(1)} &0\\
		0 & 0 & 0 & d_1^{(1)}
	\end{pmatrix},
 \label{eq:dens_dual_railv2}
\end{align}
with the coefficients $\{c_k^{l}\}$ (for the single rail state) and the $\{d_k^{l}\}$ (for the dual rail state) defined previously.

\subsection{Evaluation of State Fidelity and Hashing Bound}
Let us consider the generalized state description for both single and dual rail cases, which is a QM joint density matrix of the form
\begin{align}
	{\sigma}_{AB}=\frac{1}{\mathbf{N}} \begin{pmatrix}
		\mathbf{a}&  0 & 0 & 0 \\
		0 &\mathbf{b}  & \mathbf{c}  & 0 \\
		0 &  \mathbf{c}^*  &  \mathbf{d}&0 \\
		0 & 0 & 0 & \mathbf{e}\\
	\end{pmatrix}
\end{align}
For this subsection alone we consider $ \mathbf{a}, \mathbf{b}, \mathbf{d}, \mathbf{e}\in\mathbb{R}$ and $ \mathbf{c}\in \mathbb{C} $ with $ \mathbf{N}=\mathbf{a}+\mathbf{b}+\mathbf{d}+\mathbf{e} $. The partial trace $ \Tr_A(\hat{\sigma}_{AB}) $ yields the density matrix 
\begin{align}
	{\sigma}_{B}=\frac{1}{\mathbf{N}} \begin{pmatrix}
		\mathbf{a} +\mathbf{b}&  0  \\
		0 &\mathbf{d}+\mathbf{e}  
	\end{pmatrix}
\end{align}

The generalized expressions for the state fidelity is given as 
\begin{align}
	F({\sigma}_{AB},\ket{\Psi^\pm})=\frac{ \mathbf{b}+\mathbf{d}\pm2|\mathbf{c}|}{2\mathbf{N}}
\end{align}

For the distillable entanglement, we calculate the eigenvalues for $ {\sigma}_{AB} $ and $ \sigma_{B} $,
\begin{subequations}
	\begin{align}
		\mathrm{eig}[	{\sigma}_{AB}]&= \left\{ \frac{\mathbf{a}}{\mathbf{N}},  \frac{\mathbf{e}}{\mathbf{N}}, \frac{\mathbf{b}+\mathbf{d}\pm\sqrt{(\mathbf{b}-\mathbf{d})^2 + 4|\mathbf{c}|^2}}{\mathbf{N}}\right\} ; \\
		\mathrm{eig}[	{\sigma}_{B}]&= \left\{ \frac{\mathbf{a+b}}{\mathbf{N}},  \frac{\mathbf{d+e}}{\mathbf{N}} \right\} 
	\end{align}
\end{subequations}

\begin{table}[h!]
	\begin{tabular}{P{5cm} P{9cm} P{4cm}}
		\toprule
		\textbf{Case} & \textbf{Single Rail} & \textbf{Dual Rail}\\\midrule
        Symmetric noiseless perfectly mode-matched link $ (\eta_A=\eta_B=\sqrt{\eta}, P_d=0, \mathcal{V}=1)$ & $h_2(\frac{\gamma/2}{1-(1-\gamma)\sqrt{\eta}})- h_2(\frac{\gamma}{1-(1-\gamma)\sqrt{\eta}})$ & 1 \\ \midrule
		Asymmetric noiseless perfectly mode-matched link $ (\eta_A\neq\eta_B, P_d=0, \mathcal{V}=1)$ & $h_2\left(\frac{\eta_B(1-\gamma_B)\gamma_A}{\eta_A(1-\gamma_B)+\eta_B(1-\gamma_A) -2\eta_A\eta_B(1-\gamma_A)(1-\gamma_B)}\right)- h_2\left(\frac{\eta_A (1-\gamma_A)\gamma_B+\eta_B(1-\gamma_B)\gamma_A}{\eta_A(1-\gamma_B)+\eta_B(1-\gamma_A) -2\eta_A\eta_B(1-\gamma_A)(1-\gamma_B)}\right)$ & 1 \\ \midrule
         Symmetric noiseless  link $ (\eta_A=\eta_B=\sqrt{\eta}, P_d=0)$ & $h_2(\frac{\gamma/2}{1-(1-\gamma)\sqrt{\eta}})+\frac{\gamma(1+\mathcal{V})/2}{1-(1-\gamma)\sqrt{\eta}}\log_2\left(\frac{\gamma(1+\mathcal{V})/2}{1-(1-\gamma)\sqrt{\eta}}\right)
         +\frac{\gamma(1-\mathcal{V})/2}{1-(1-\gamma)\sqrt{\eta}}\log_2\left(\frac{\gamma(1-\mathcal{V})/2}{1-(1-\gamma)\sqrt{\eta}}\right) + \left(1-\frac{\gamma}{1-(1-\gamma)\sqrt{\eta}}\right) \log_2 \left(1-\frac{\gamma}{1-(1-\gamma)\sqrt{\eta}}\right)$ & $1-h_2(\frac{1-\mathcal{V}^2}{2})$ \\ 
		\bottomrule
	\end{tabular}
	\caption{\label{tab:RCI_spl} Formula for $I(\rho_{AB})$ for a few special cases. }
\end{table}

\begin{table}[h!]
	\begin{tabular}{P{5cm} P{9cm} P{4cm}}
		\toprule
		\textbf{Case} & \textbf{Single Rail} & \textbf{Dual Rail}\\\midrule
        Symmetric noiseless perfectly mode-matched link $ (\eta_A=\eta_B=\sqrt{\eta}, P_d=0, \mathcal{V}=1)$ & $\frac{\gamma}{1-(1-\gamma)\sqrt{\eta}}$ & 1 \\ \midrule
		Asymmetric noiseless perfectly mode-matched link $ (\eta_A\neq\eta_B, P_d=0, \mathcal{V}=1)$ & $\frac{\gamma _B\left(1-\gamma_A\right) \eta_A +\gamma_A \left(1-\gamma _B\right) \eta _B + 2 \sqrt{\eta_A\eta_B} \sqrt{\gamma_A\gamma_B\left(1-\gamma _A\right)\left(1-\gamma _B\right)}}{2
   \left(1-\gamma _B\right) \eta _B + 2 \left(1-\gamma _A\right) \eta _A  + 4\eta_A\eta_B(1-\gamma_A)(1-\gamma_B)}$ & 1 \\ \midrule
         Symmetric noiseless  link $ (\eta_A=\eta_B=\sqrt{\eta}, P_d=0)$ & $\frac{(1+\mathcal{V})\gamma/2}{1-(1-\gamma)\sqrt{\eta}}$ & $(1+\mathcal{V}^2)/2$ \\ 
		\bottomrule
	\end{tabular}
	\caption{\label{tab:fidelity_spl} Formula for $F(\rho_{AB},\ket{\Psi^{\pm}})$ for a few special cases. }
\end{table}

\subsection{Analysis of Asymmetric Links ($\eta_A\neq\eta_B$)}

\begin{figure}[ht]
	\centering
	\includegraphics[width=0.8\linewidth]{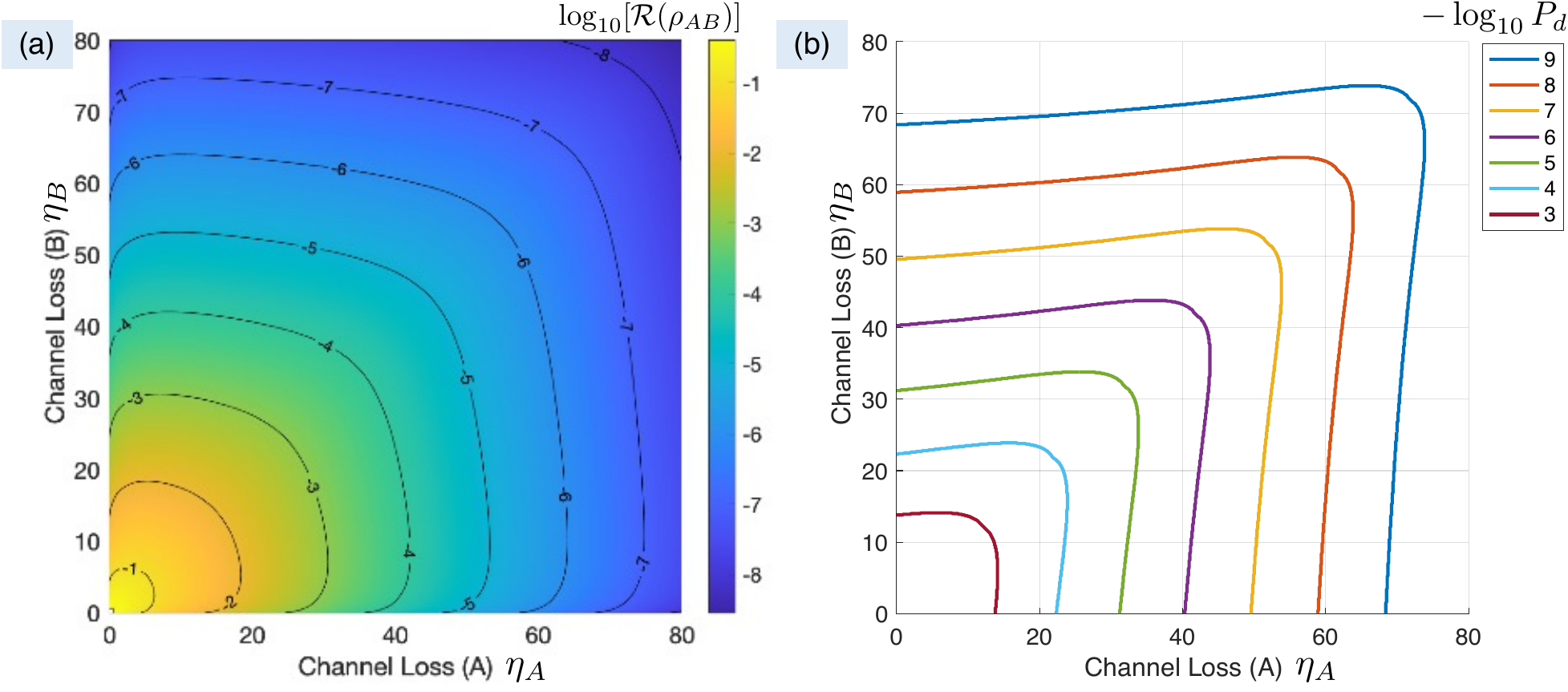}
	\caption{ (a) Rate of entanglement generation (color axis; $\log_{10}$ scale) for a single rail swap for general $\eta_A$ (horizontal axis) and $\eta_B$ (vertical axis) with optimal qubit parameters $\gamma_A$, $\gamma_B$ chosen. We assume no excess noise ($P_d=0$) and perfect mode matching ($\mathcal{V}=1$) with no carrier phase mismatch ($\varepsilon=0$). (b) Boundary of maximum network operation with excess noise (colored lines) for which $\mathcal{R}(\rho_{AB})>0$ for general $\eta_A$ (horizontal axis) and $\eta_B$ (vertical axis).  }
	\label{fig:sr_swaps1}
\end{figure}

\begin{figure}[ht]
	\centering
	\includegraphics[width=0.8\linewidth]{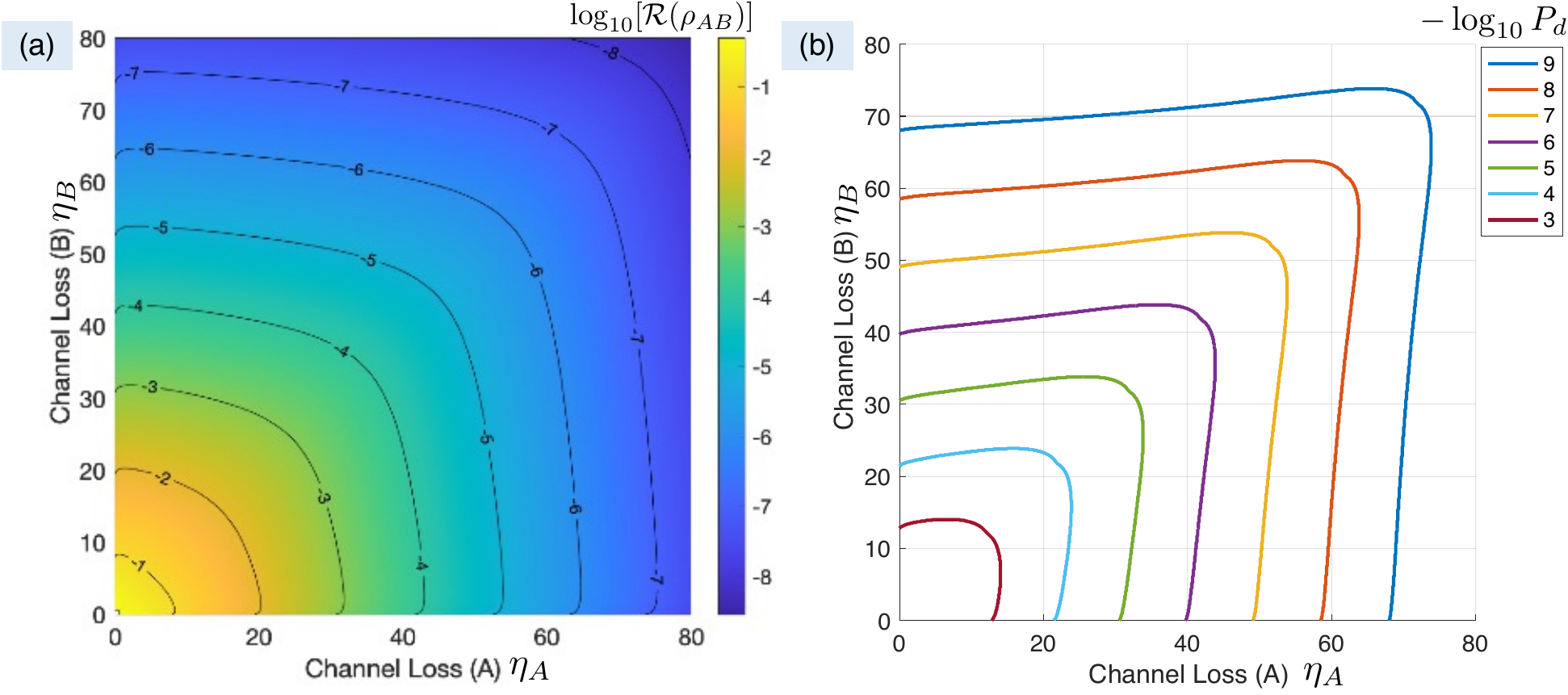}
	\caption{ (a) Rate of entanglement generation (color axis; $\log_{10}$ scale) for a single rail swap for general $\eta_A$ (horizontal axis) and $\eta_B$ (vertical axis) with equal qubit parameters $\gamma_A=\gamma_B$ chosen. We assume no excess noise ($P_d=0$) and perfect mode matching ($\mathcal{V}=1$) with no carrier phase mismatch ($\varepsilon=0$). (b) Boundary of maximum network operation with excess noise (colored lines) for which $\mathcal{R}(\rho_{AB})>0$ for general $\eta_A$ (horizontal axis) and $\eta_B$ (vertical axis).}
	\label{fig:sr_swaps2}
\end{figure}

\begin{figure}[ht]
	\centering
	\includegraphics[width=0.8\linewidth]{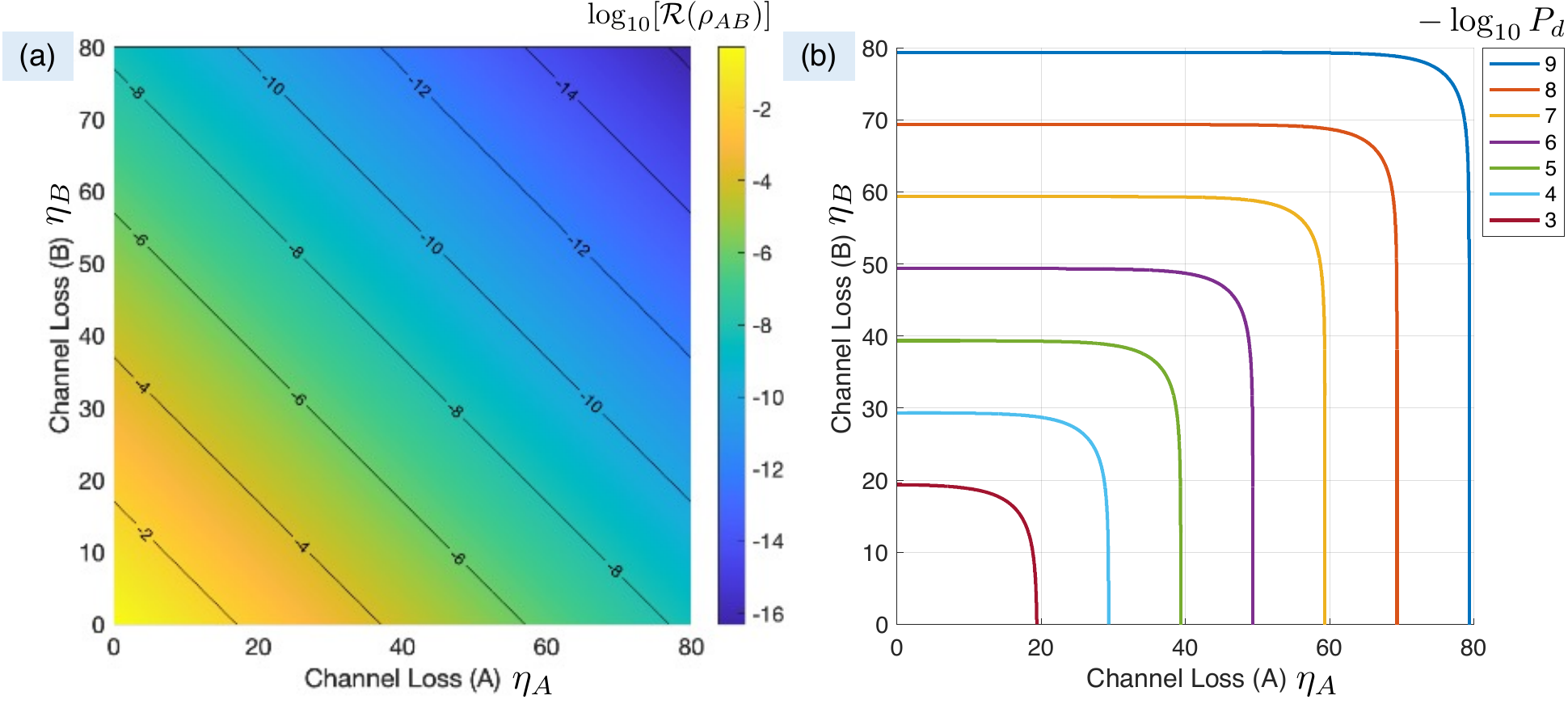}
	\caption{(a) Rate of entanglement generation (color axis; $\log_{10}$ scale) for a dual rail swap for general $\eta_A$ (x-axis) and $\eta_B$ (y-axis), assuming no excess noise ($P_d=0$) and perfect mode matching ($\mathcal{V}=1$). (b) Boundary of maximum network operation with excess noise (colored lines) for which $\mathcal{R}(\rho_{AB})>0$ for general $\eta_A$ (x-axis) and $\eta_B$ (y-axis). }
	\label{fig:dr_swaps}
\end{figure}

In the main text we have limited the discussion of the effect of various hardware parameters to the state quality to the case of symmetric links ($\eta_A=\eta_B=\sqrt{\eta}$). In this Appendix we present the case for general links which allows for asymmetry in the channel transmissivities. Figs.~\ref{fig:sr_swaps1}-\ref{fig:dr_swaps} (a) show the rate plots for the single rail and dual rail swaps with general $\eta_A$ and $\eta_B$ assuming no excess noise ($P_d=0$) and perfect mode matching ($\mathcal{V}=1$) with no carrier phase mismatch ($\varepsilon=0$). 

In Figs.~\ref{fig:sr_swaps1}-\ref{fig:dr_swaps}(b), we evaluate the max range of link efficiencies, $(\eta_A,\eta_B)$ for which entanglement swapping yields $\mathcal{R}(\rho_{AB})>0$ in the presence of excess noise ($P_d>0$). This denotes the region over which the swapping protocols yields bipartite states with non-zero distillable entanglement.

        \section{ Carrier Phase Mismatch from Atmospheric Turbulence}
\label{app:phase_mismatch}

The effect of carrier phase mismatch is to impart a a complex phase to the mode mismatch parameter $\mathcal{V}$, i.e., a purely real $\mathcal{V}$ becomes a complex number where $|\mathcal{V}|$ is the original real mode-mismatch parameter and $\arg \mathcal{V}=\theta$, which is a random variable drawn from a distribution. For free-space optical (FSO) links, $\theta$ is sampled from a zero-mean Gaussian distribution, $\mathcal{N}(0,\sigma_\phi^2)$ whose variance, $\sigma_\phi^2$ is determined by the strength of atmospheric turbulence. Considering the Kolmogorov spectrum of turbulence~\cite{Andrews2005-zi} for quasi-monochromatic radiation, the variance of the random phase distribution~\cite{Belmonte2009-hb} can be determined by
\begin{align}
    \sigma^2_\phi=C_J \left(\frac{d_R}{r_0}\right)^{5/3}
    \label{eq:phvar}
\end{align}
where $d_R$ is the diameter of the receiver aperture in the link, $r_0$ is the Fried parameter and $C_f$ is the coefficient that determines the order of correction for the aberrations. The Fried parameter is given by 
\begin{align}
    r_0=1.68 C^2_n L\left(\frac{2\pi}{\lambda}\right)^2
\end{align}
where $\lambda$ is the wavelength of the transmitted light, $L$ is the propagation distance and $C^2_n$ is the atmospheric structure constant, which is a measure of the strength of the turbulence induced scintillation.

For systems with no phase correction (i.e.\ direct transmission with no adaptive optics), $C_J=1.0299$ is used in Eq.~\eqref{eq:phvar}~\cite{Belmonte2009-hb,Noll1976-aw}. The standard method for systems employing adaptive optics is to correct for phase aberration terms upto the corresponding $J$-th order Zernike polynomial expansion -- the coefficient for these corrections can been pre-calculated~\cite{Noll1976-aw}. For large $J$, we may use the approximate formula $C_J\approx 0.2944J^{-\sqrt{3}/2}$. 

\begin{figure}[h!]
	\centering
	\includegraphics[width=0.4\linewidth]{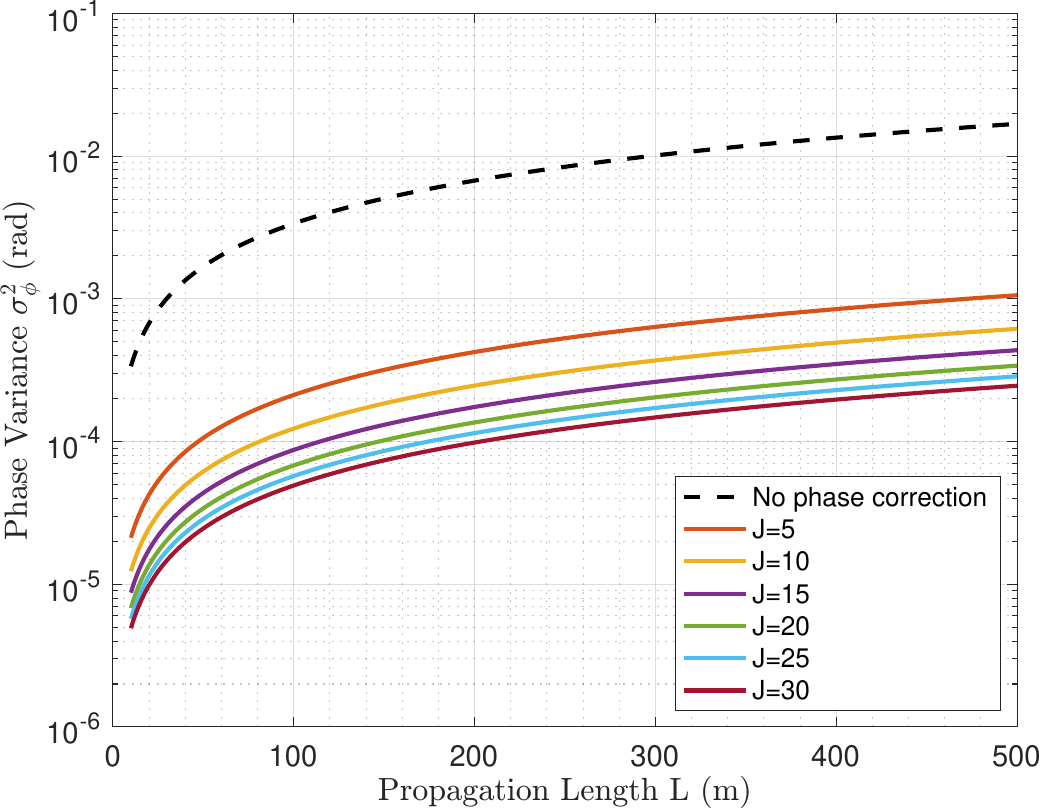}
	\caption{Phase variance induced by atmospheric turbulence over FSO links with $L\in[10,500]$ m. We assume transmission at $\lambda=1550$ nm, $C^2_n=1.5\times 10^{-17}$ and $d_R=0.5$ m. The black dashed line indicates the performance with phase correction $(C_J=1.0299)$, we consider correction (colored lines) for Zernicke polynomial expansion upto $J\in\{5,10,15,20,25\}$, with corresponding $C_J=\{0.0648,0.0377,0.0267,0.0208,0.0181,0.0155\}$. }
	\label{fig:phase_varv2}
\end{figure}

\begin{figure}[h!]
	\centering
	\includegraphics[width=0.4\linewidth]{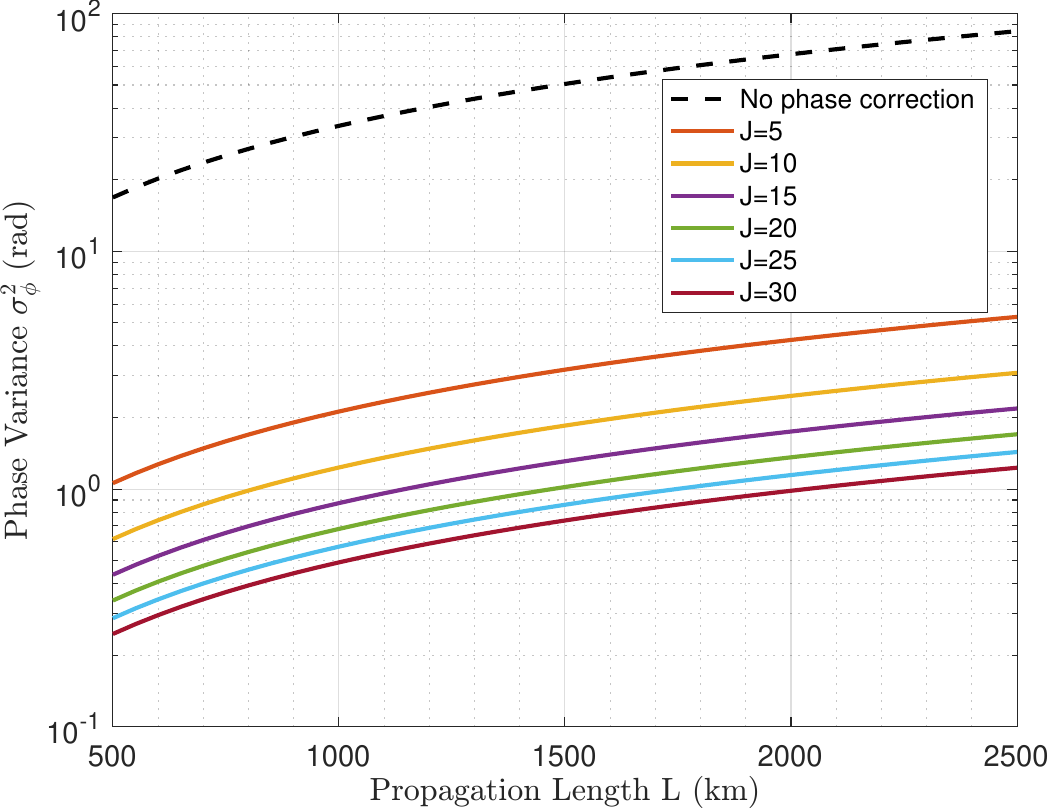}
	\caption{Phase variance induced by atmospheric turbulence over FSO links with $L\in[500,2500]$ km. We assume transmission at $\lambda=1550$ nm, $C^2_n=1.5\times 10^{-17}$ and $d_R=0.5$ m. The black dashed line indicates the performance with phase correction $(C_J=1.0299)$, we consider correction (colored lines) for Zernicke polynomial expansion upto $J\in\{5,10,15,20,25\}$, with corresponding $C_J=\{0.0648,0.0377,0.0267,0.0208,0.0181,0.0155\}$.}
	\label{fig:phase_varv1}
\end{figure}

We plot the value of phase variance (in radians) for FSO lengths in the range of 10-500 m (representing terrestrial FSO links) in Fig.~\ref{fig:phase_varv2} and 500-2500 km (representing space-to-ground FSO links for low and medium-earth orbit satellites) in Fig.~\ref{fig:phase_varv1}. We assume transmission at $\lambda=1550$ nm with atmospheric structure constant $C^2_n=1.5\times 10^{-17}$ coupling into a receiver of diameter $d_R=0.5$ m, with pre-calculated values for $C_J$ (indicated in the figure caption).

	\section{Analysis of Distillation Circuit }
\label{app:distill}

\subsection{Iterative Map for Analytical Proofs }
\begin{figure}[h!]
	\centering
	\includegraphics[width=0.8\linewidth]{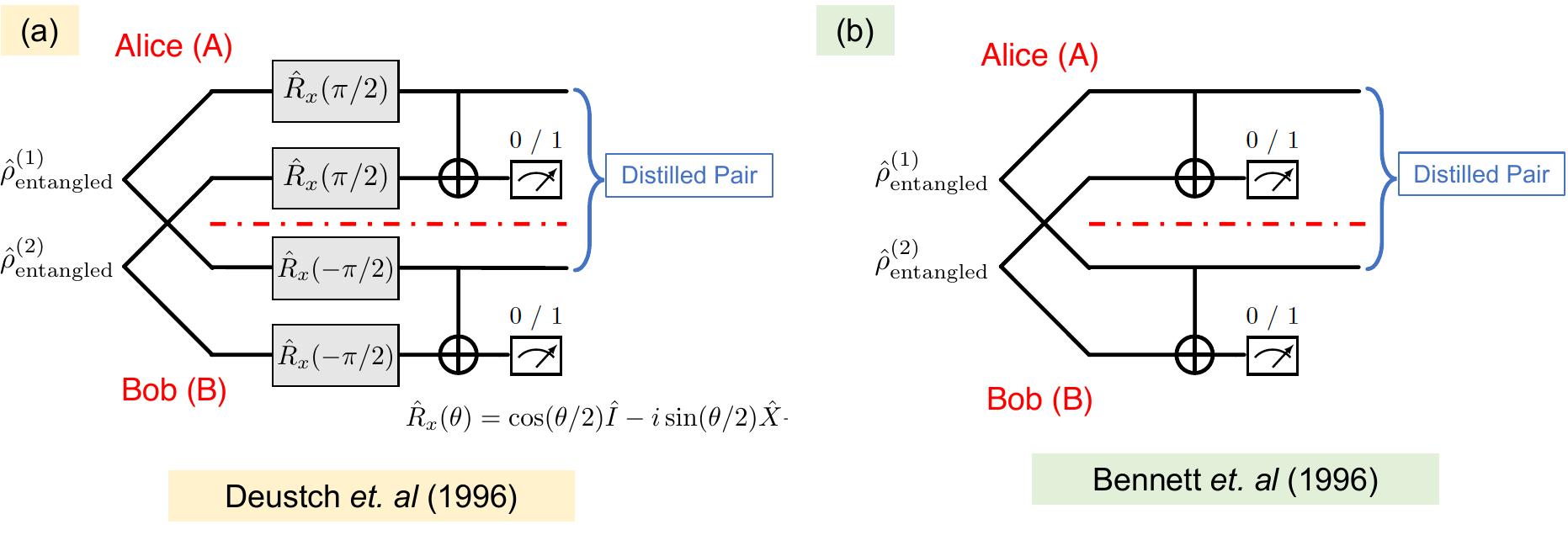}
	\caption{ Circuits that implement 2-to-1 entanglement distillation (a) proposed by Deustch \emph{et. al} in~\cite{Deutsch1996-sb}, and, (b) proposed by Bennett \emph{et. al} in~\cite{Bennett1996-vn}.  }
	\label{fig:dist_comp}
\end{figure}

We consider the use of distillation circuits for the improvement of the heralded state quality and entanglement distribution utility in the presence of network non-idealities. Specifically we use the protocol proposed in~\cite{Deutsch1996-sb}, illustrated by Fig.~\ref{fig:dist_comp}(a). Starting with the general states (as shown in Appendix~\ref{app:general_state}) and applying the distillation circuit is cumbersome to analyze analytically. However we may consider the specific non-ideal case of the dual rail swap with no excess noise $ P_d=0 $, with $ \mathcal{V}<1 $ and show the utility of distillation. 

As analyzed in~\cite{Deutsch1996-sb}, we start with a mixed entangled state which is Bell-diagonal and of the form,
\begin{align}	\hat{\rho}=A\outprod{\Psi^+}+B\outprod{\Psi^-}+C\outprod{\Phi^+}+D\outprod{\Phi^-}.
\end{align}
Such a state maybe compactly represented as a vector $ \left[A,B,C,D\right] $. The action of the Deutsch \emph{et. al} distillation circuit also yields a Bell-diagonal state represented in the vector format as $ [A',B',C',D'] $ where the following transformation rules hold,
\begin{align}
\begin{split}
		A'&=(A^2+D^2)/N\\
	B' &= 2AD/N\\
	C'&=B^2+C^2/N\\
	D'&=2BC/N,
	\label{eq:map_rule}
\end{split}
\end{align}
with $ N=(A+D)^2+(B+C)^2 $. Returning to our special case ($ P_d=0; \mathcal{V}<1 $), with $\mathcal{V}'=|\mathcal{V}|^2$ we may express the dual-rail heralded state as $ 	\hat{\rho}_{d}=(({1+\mathcal{V}'}) \outprod{\Psi^+ }+({1-\mathcal{V}'}) \outprod{\Psi^-} )/2 $. Hence as per our vector notation
\begin{align}
	\text{Original state:} \begin{cases}
	 	A=(1+\mathcal{V}')/2 \\
	 	B=(1-\mathcal{V}')/2 \\
	 	 C=D=0 .
	 \end{cases} 
\end{align}
A single round of distillation yields the state (in vector notation),
\begin{align}
	\text{One round: }\begin{cases}
		A'={A^2}/{(A^2+B^2)}\\
		C'={B^2}/{(A^2+B^2)}\\
		B'=D'=0.
	\end{cases}
\end{align}
Since $ B'=D'=0 $, for subsequent rounds these components \emph{always} remain zero as per the mapping rule. Consequently after $ k $-rounds of distillation, the $ A $ component (say represented by $ A^{(k)} $) is given as $A^{(k)}:=A^{2^k}/(A^{2^k}+B^{2^k}) $. This dictates the fidelity of the distilled state w.r.t.\ the target $ \ket{\Psi^+} $ state. We may rearrange terms to represent $ A^{(k)} $ as 
\begin{subequations}
	\begin{align}
		A^{(k)}=\frac{(	1+\mathcal{V}')^{K}}{(	1+\mathcal{V}')^{K}+(	1-\mathcal{V}')^{K}} \label{eq:ineq1}\\
		=\left( {1+\left(\frac{1-\mathcal{V}'}{1+\mathcal{V}'}\right)^K}\right)^{-1},
	\end{align}
\end{subequations}
where $ K=2^k $. Since $ \mathcal{V}'\in[0,1] $, $ (1-\mathcal{V}')/(1+\mathcal{V}')<1 $, this implies the denominator of Eq.~\eqref{eq:ineq1} decreases as $ k $ increases. Consequently $ A^{(k+1)} \geq A^{(k)}$, with equality holding for $ \mathcal{V}'=1 $, this implies the fidelity of the state increases as $ k $ (i.e. the rounds of distillation) increases. One may equivalently analyze the hashing bound for this hierarchy of states to prove the viability of using the distillation circuit.  

In contrast, applying the alternate distillation circuit (see Fig.~\ref{fig:dist_comp}b) introduced by Bennett \emph{et. al} in~\cite{Bennett1996-vn} shows no advantage. Starting with a pre-distillation mixed state which is Bell diagonal and representable in the vector format, $ [A,B,C,D] $, the state after distillation is also Bell diagonal with the underlying map,
\begin{align}
	\begin{split}
	A'&=(A^2+B^2)/2\\
	B' &= AB\\
	C'&=(C^2+D^2)/2\\
	D'&=CD
	\label{eq:map_rule2}
\end{split}
\end{align}
Starting out with a dual rail swap heralded state with no excess noise ($ P_d=0; \mathcal{V}<1 $) as before, 
\begin{align}
	\text{Original state:} \begin{cases}
		A=(1+\mathcal{V}^2)/2 \\
		B=(1-\mathcal{V}^2)/2 \\
		C=D=0, 
	\end{cases} 
\end{align} which gives
\begin{align}
	\text{One round:} \begin{cases}
		A'=(1+\mathcal{V}^4)/2 \\
		B=(1-\mathcal{V}^4)/2 \\
		C=D=0 
	\end{cases} 
\end{align}
Similarly after $ k $-rounds of distillation fidelity is given by  $ A^{(k)} =(1+\mathcal{V}^{2^{k+1}})/2$. Since $ \mathcal{V}\in [0,1] $, this number decreases with $ k $, which means that additional rounds of this protocol is not beneficial to improving the state quality.

\subsection{Max Network Range}

\begin{figure}[h!]
	\centering
	\includegraphics[width=0.8\linewidth]{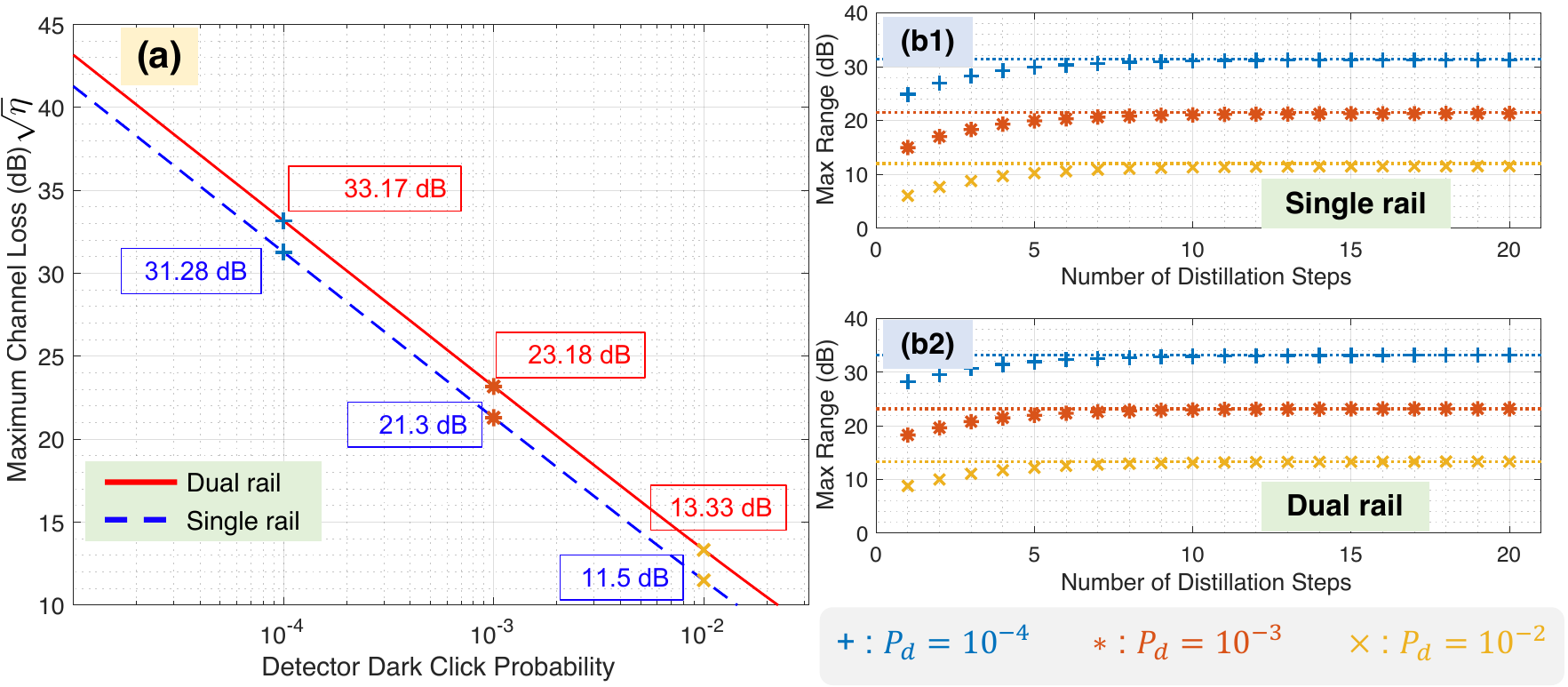}
	\caption{ Maximum range of using distillation circuits $ \eta_{\lim} $(a) using analytically extracted values by solving $ F(\rho_{AB},\ket{\Psi^+})=0.5 $ for $ \eta $ at given values of $ P_d $ and $ \mathcal{V} $, and shown (b) using evaluation of distilled states and numerical extraction of $ \eta_{\lim} $. Specific point values are extracted for $ P_d= \{10^{-4},10^{-3},10^{-2}\}$ for both single (blue) and dual rail (red) swaps. }
	\label{fig:dist_sat}
\end{figure}

In the main text we have highlighted how the max range (i.e.\ value of $ \eta $ at which $ I(\rho) \rightarrow0$) increases with each subsequent round of distillation and saturates to a limiting value.  We label this as $ \eta_{\lim} $ i.e.\ the channel loss where $ F(\rho_{AB},\ket{\Psi^+})=0.5 $. The straightforward reason is that at  $ F(\rho_{AB},\ket{\Psi^+})=0.5 $ the initial joint state is the two-qubit maximally mixed state $ \rho_{AB}=\mathbb{I}/4 $ with no distillable entanglement i.e.\ $ E_D(\mathbb{I}/4)=0 $. Hence no amount of distillation would improve the state quality. For all $ \eta<\eta_{\lim} $, the heralded state satisfies $ F(\rho_{AB},\ket{\Psi^+})>0.5  $ leaving room for improvement of state quality by repeated distillation, allowing $ I(\rho) $ to be improved asymptotically to 1. 

Since a complete analytical treatment for general $ P_d >0 $  is intractable, we show some numerical results in Fig.~\ref{fig:dist_sat} that highlights the result. We solve $ F(\rho_{AB},\ket{\Psi^+})=0.5 $ to extract $ \eta_{\lim} $ for a range of $ P_d \in[10^{-4},10^{-1}] $ at $ \mathcal{V}=1 $ and plot the same in Fig.~\ref{fig:dist_sat}(a). Values of $ \eta_{\lim} $ for $ P_d= \{10^{-4},10^{-3},10^{-2}\}$  (corresponding symbols in legend) are marked for the single rail (blue text) and dual rail (red text). We then look at the max range for the corresponding values of $ P_d $ vs. rounds of distillation for the single rail Fig.~\ref{fig:dist_sat}(b1) and dual rail Fig.~\ref{fig:dist_sat}(b2) swap. The calculated and  numerically extracted $ \eta_{\lim} $ values (marked by lines) show close agreement.

	\twocolumngrid
	
	\bibliography{biblio_singvsdual.bib}

\begin{thebibliography}{31}%
\makeatletter
\providecommand \@ifxundefined [1]{%
 \@ifx{#1\undefined}
}%
\providecommand \@ifnum [1]{%
 \ifnum #1\expandafter \@firstoftwo
 \else \expandafter \@secondoftwo
 \fi
}%
\providecommand \@ifx [1]{%
 \ifx #1\expandafter \@firstoftwo
 \else \expandafter \@secondoftwo
 \fi
}%
\providecommand \natexlab [1]{#1}%
\providecommand \enquote  [1]{``#1''}%
\providecommand \bibnamefont  [1]{#1}%
\providecommand \bibfnamefont [1]{#1}%
\providecommand \citenamefont [1]{#1}%
\providecommand \href@noop [0]{\@secondoftwo}%
\providecommand \href [0]{\begingroup \@sanitize@url \@href}%
\providecommand \@href[1]{\@@startlink{#1}\@@href}%
\providecommand \@@href[1]{\endgroup#1\@@endlink}%
\providecommand \@sanitize@url [0]{\catcode `\\12\catcode `\$12\catcode
  `\&12\catcode `\#12\catcode `\^12\catcode `\_12\catcode `\%12\relax}%
\providecommand \@@startlink[1]{}%
\providecommand \@@endlink[0]{}%
\providecommand \url  [0]{\begingroup\@sanitize@url \@url }%
\providecommand \@url [1]{\endgroup\@href {#1}{\urlprefix }}%
\providecommand \urlprefix  [0]{URL }%
\providecommand \Eprint [0]{\href }%
\providecommand \doibase [0]{https://doi.org/}%
\providecommand \selectlanguage [0]{\@gobble}%
\providecommand \bibinfo  [0]{\@secondoftwo}%
\providecommand \bibfield  [0]{\@secondoftwo}%
\providecommand \translation [1]{[#1]}%
\providecommand \BibitemOpen [0]{}%
\providecommand \bibitemStop [0]{}%
\providecommand \bibitemNoStop [0]{.\EOS\space}%
\providecommand \EOS [0]{\spacefactor3000\relax}%
\providecommand \BibitemShut  [1]{\csname bibitem#1\endcsname}%
\let\auto@bib@innerbib\@empty
\bibitem [{\citenamefont {Barrett}\ and\ \citenamefont
  {Kok}(2005)}]{Barrett2004-sj}%
  \BibitemOpen
  \bibfield  {author} {\bibinfo {author} {\bibfnamefont {S.~D.}\ \bibnamefont
  {Barrett}}\ and\ \bibinfo {author} {\bibfnamefont {P.}~\bibnamefont {Kok}},\
  }\bibfield  {title} {\bibinfo {title} {Efficient high-fidelity quantum
  computation using matter qubits and linear optics},\ }\href
  {https://link.aps.org/doi/10.1103/PhysRevA.71.060310} {\bibfield  {journal}
  {\bibinfo  {journal} {Phys. Rev. A}\ }\textbf {\bibinfo {volume} {71}},\
  \bibinfo {pages} {060310} (\bibinfo {year} {2005})}\BibitemShut {NoStop}%
\bibitem [{\citenamefont {Kalb}\ \emph {et~al.}(2017)\citenamefont {Kalb},
  \citenamefont {Reiserer}, \citenamefont {Humphreys}, \citenamefont
  {Bakermans}, \citenamefont {Kamerling}, \citenamefont {Nickerson},
  \citenamefont {Benjamin}, \citenamefont {Twitchen}, \citenamefont {Markham},\
  and\ \citenamefont {Hanson}}]{Kalb2017-hr}%
  \BibitemOpen
  \bibfield  {author} {\bibinfo {author} {\bibfnamefont {N.}~\bibnamefont
  {Kalb}}, \bibinfo {author} {\bibfnamefont {A.~A.}\ \bibnamefont {Reiserer}},
  \bibinfo {author} {\bibfnamefont {P.~C.}\ \bibnamefont {Humphreys}}, \bibinfo
  {author} {\bibfnamefont {J.~J.~W.}\ \bibnamefont {Bakermans}}, \bibinfo
  {author} {\bibfnamefont {S.~J.}\ \bibnamefont {Kamerling}}, \bibinfo {author}
  {\bibfnamefont {N.~H.}\ \bibnamefont {Nickerson}}, \bibinfo {author}
  {\bibfnamefont {S.~C.}\ \bibnamefont {Benjamin}}, \bibinfo {author}
  {\bibfnamefont {D.~J.}\ \bibnamefont {Twitchen}}, \bibinfo {author}
  {\bibfnamefont {M.}~\bibnamefont {Markham}},\ and\ \bibinfo {author}
  {\bibfnamefont {R.}~\bibnamefont {Hanson}},\ }\bibfield  {title} {\bibinfo
  {title} {Entanglement distillation between solid-state quantum network
  nodes},\ }\href {http://dx.doi.org/10.1126/science.aan0070} {\bibfield
  {journal} {\bibinfo  {journal} {Science}\ }\textbf {\bibinfo {volume}
  {356}},\ \bibinfo {pages} {928} (\bibinfo {year} {2017})}\BibitemShut
  {NoStop}%
\bibitem [{\citenamefont {Pompili}\ \emph {et~al.}(2021)\citenamefont
  {Pompili}, \citenamefont {Hermans}, \citenamefont {Baier}, \citenamefont
  {Beukers}, \citenamefont {Humphreys}, \citenamefont {Schouten}, \citenamefont
  {Vermeulen}, \citenamefont {Tiggelman}, \citenamefont {Dos Santos~Martins},
  \citenamefont {Dirkse}, \citenamefont {Wehner},\ and\ \citenamefont
  {Hanson}}]{Pompili2021-bp}%
  \BibitemOpen
  \bibfield  {author} {\bibinfo {author} {\bibfnamefont {M.}~\bibnamefont
  {Pompili}}, \bibinfo {author} {\bibfnamefont {S.~L.~N.}\ \bibnamefont
  {Hermans}}, \bibinfo {author} {\bibfnamefont {S.}~\bibnamefont {Baier}},
  \bibinfo {author} {\bibfnamefont {H.~K.~C.}\ \bibnamefont {Beukers}},
  \bibinfo {author} {\bibfnamefont {P.~C.}\ \bibnamefont {Humphreys}}, \bibinfo
  {author} {\bibfnamefont {R.~N.}\ \bibnamefont {Schouten}}, \bibinfo {author}
  {\bibfnamefont {R.~F.~L.}\ \bibnamefont {Vermeulen}}, \bibinfo {author}
  {\bibfnamefont {M.~J.}\ \bibnamefont {Tiggelman}}, \bibinfo {author}
  {\bibfnamefont {L.}~\bibnamefont {Dos Santos~Martins}}, \bibinfo {author}
  {\bibfnamefont {B.}~\bibnamefont {Dirkse}}, \bibinfo {author} {\bibfnamefont
  {S.}~\bibnamefont {Wehner}},\ and\ \bibinfo {author} {\bibfnamefont
  {R.}~\bibnamefont {Hanson}},\ }\bibfield  {title} {\bibinfo {title}
  {Realization of a multinode quantum network of remote solid-state qubits},\
  }\href {https://www.science.org/doi/10.1126/science.abg1919} {\bibfield
  {journal} {\bibinfo  {journal} {Science}\ }\textbf {\bibinfo {volume}
  {372}},\ \bibinfo {pages} {259} (\bibinfo {year} {2021})}\BibitemShut
  {NoStop}%
\bibitem [{\citenamefont {Hermans}\ \emph {et~al.}(2022)\citenamefont
  {Hermans}, \citenamefont {Pompili}, \citenamefont {Beukers}, \citenamefont
  {Baier}, \citenamefont {Borregaard},\ and\ \citenamefont
  {Hanson}}]{Hermans2022-zl}%
  \BibitemOpen
  \bibfield  {author} {\bibinfo {author} {\bibfnamefont {S.~L.~N.}\
  \bibnamefont {Hermans}}, \bibinfo {author} {\bibfnamefont {M.}~\bibnamefont
  {Pompili}}, \bibinfo {author} {\bibfnamefont {H.~K.~C.}\ \bibnamefont
  {Beukers}}, \bibinfo {author} {\bibfnamefont {S.}~\bibnamefont {Baier}},
  \bibinfo {author} {\bibfnamefont {J.}~\bibnamefont {Borregaard}},\ and\
  \bibinfo {author} {\bibfnamefont {R.}~\bibnamefont {Hanson}},\ }\bibfield
  {title} {\bibinfo {title} {Qubit teleportation between non-neighbouring nodes
  in a quantum network},\ }\href {http://dx.doi.org/10.1038/s41586-022-04697-y}
  {\bibfield  {journal} {\bibinfo  {journal} {Nature}\ }\textbf {\bibinfo
  {volume} {605}},\ \bibinfo {pages} {663} (\bibinfo {year}
  {2022})}\BibitemShut {NoStop}%
\bibitem [{\citenamefont {Gao}\ \emph {et~al.}(2012)\citenamefont {Gao},
  \citenamefont {Fallahi}, \citenamefont {Togan}, \citenamefont
  {Miguel-Sanchez},\ and\ \citenamefont {Imamoglu}}]{Gao2012-eh}%
  \BibitemOpen
  \bibfield  {author} {\bibinfo {author} {\bibfnamefont {W.~B.}\ \bibnamefont
  {Gao}}, \bibinfo {author} {\bibfnamefont {P.}~\bibnamefont {Fallahi}},
  \bibinfo {author} {\bibfnamefont {E.}~\bibnamefont {Togan}}, \bibinfo
  {author} {\bibfnamefont {J.}~\bibnamefont {Miguel-Sanchez}},\ and\ \bibinfo
  {author} {\bibfnamefont {A.}~\bibnamefont {Imamoglu}},\ }\bibfield  {title}
  {\bibinfo {title} {Observation of entanglement between a quantum dot spin and
  a single photon},\ }\href {http://dx.doi.org/10.1038/nature11573} {\bibfield
  {journal} {\bibinfo  {journal} {Nature}\ }\textbf {\bibinfo {volume} {491}},\
  \bibinfo {pages} {426} (\bibinfo {year} {2012})}\BibitemShut {NoStop}%
\bibitem [{\citenamefont {Hucul}\ \emph {et~al.}(2014)\citenamefont {Hucul},
  \citenamefont {Inlek}, \citenamefont {Vittorini}, \citenamefont {Crocker},
  \citenamefont {Debnath}, \citenamefont {Clark},\ and\ \citenamefont
  {Monroe}}]{Hucul2014-jj}%
  \BibitemOpen
  \bibfield  {author} {\bibinfo {author} {\bibfnamefont {D.}~\bibnamefont
  {Hucul}}, \bibinfo {author} {\bibfnamefont {I.~V.}\ \bibnamefont {Inlek}},
  \bibinfo {author} {\bibfnamefont {G.}~\bibnamefont {Vittorini}}, \bibinfo
  {author} {\bibfnamefont {C.}~\bibnamefont {Crocker}}, \bibinfo {author}
  {\bibfnamefont {S.}~\bibnamefont {Debnath}}, \bibinfo {author} {\bibfnamefont
  {S.~M.}\ \bibnamefont {Clark}},\ and\ \bibinfo {author} {\bibfnamefont
  {C.}~\bibnamefont {Monroe}},\ }\bibfield  {title} {\bibinfo {title} {Modular
  entanglement of atomic qubits using photons and phonons},\ }\href
  {https://www.nature.com/articles/nphys3150} {\bibfield  {journal} {\bibinfo
  {journal} {Nat. Phys.}\ }\textbf {\bibinfo {volume} {11}},\ \bibinfo {pages}
  {37} (\bibinfo {year} {2014})}\BibitemShut {NoStop}%
\bibitem [{\citenamefont {Inlek}\ \emph {et~al.}(2017)\citenamefont {Inlek},
  \citenamefont {Crocker}, \citenamefont {Lichtman}, \citenamefont {Sosnova},\
  and\ \citenamefont {Monroe}}]{Inlek2017-tg}%
  \BibitemOpen
  \bibfield  {author} {\bibinfo {author} {\bibfnamefont {I.~V.}\ \bibnamefont
  {Inlek}}, \bibinfo {author} {\bibfnamefont {C.}~\bibnamefont {Crocker}},
  \bibinfo {author} {\bibfnamefont {M.}~\bibnamefont {Lichtman}}, \bibinfo
  {author} {\bibfnamefont {K.}~\bibnamefont {Sosnova}},\ and\ \bibinfo {author}
  {\bibfnamefont {C.}~\bibnamefont {Monroe}},\ }\bibfield  {title} {\bibinfo
  {title} {Multispecies {Trapped-Ion} node for quantum networking},\ }\href
  {http://dx.doi.org/10.1103/PhysRevLett.118.250502} {\bibfield  {journal}
  {\bibinfo  {journal} {Phys. Rev. Lett.}\ }\textbf {\bibinfo {volume} {118}},\
  \bibinfo {pages} {250502} (\bibinfo {year} {2017})}\BibitemShut {NoStop}%
\bibitem [{\citenamefont {Stephenson}\ \emph {et~al.}(2020)\citenamefont
  {Stephenson}, \citenamefont {Nadlinger}, \citenamefont {Nichol},
  \citenamefont {An}, \citenamefont {Drmota}, \citenamefont {Ballance},
  \citenamefont {Thirumalai}, \citenamefont {Goodwin}, \citenamefont {Lucas},\
  and\ \citenamefont {Ballance}}]{Stephenson2020-rw}%
  \BibitemOpen
  \bibfield  {author} {\bibinfo {author} {\bibfnamefont {L.~J.}\ \bibnamefont
  {Stephenson}}, \bibinfo {author} {\bibfnamefont {D.~P.}\ \bibnamefont
  {Nadlinger}}, \bibinfo {author} {\bibfnamefont {B.~C.}\ \bibnamefont
  {Nichol}}, \bibinfo {author} {\bibfnamefont {S.}~\bibnamefont {An}}, \bibinfo
  {author} {\bibfnamefont {P.}~\bibnamefont {Drmota}}, \bibinfo {author}
  {\bibfnamefont {T.~G.}\ \bibnamefont {Ballance}}, \bibinfo {author}
  {\bibfnamefont {K.}~\bibnamefont {Thirumalai}}, \bibinfo {author}
  {\bibfnamefont {J.~F.}\ \bibnamefont {Goodwin}}, \bibinfo {author}
  {\bibfnamefont {D.~M.}\ \bibnamefont {Lucas}},\ and\ \bibinfo {author}
  {\bibfnamefont {C.~J.}\ \bibnamefont {Ballance}},\ }\bibfield  {title}
  {\bibinfo {title} {{High-Rate}, {High-Fidelity} entanglement of qubits across
  an elementary quantum network},\ }\href
  {http://dx.doi.org/10.1103/PhysRevLett.124.110501} {\bibfield  {journal}
  {\bibinfo  {journal} {Phys. Rev. Lett.}\ }\textbf {\bibinfo {volume} {124}},\
  \bibinfo {pages} {110501} (\bibinfo {year} {2020})}\BibitemShut {NoStop}%
\bibitem [{\citenamefont {Krutyanskiy}\ \emph {et~al.}(2023)\citenamefont
  {Krutyanskiy}, \citenamefont {Galli}, \citenamefont {Krcmarsky},
  \citenamefont {Baier}, \citenamefont {Fioretto}, \citenamefont {Pu},
  \citenamefont {Mazloom}, \citenamefont {Sekatski}, \citenamefont {Canteri},
  \citenamefont {Teller}, \citenamefont {Schupp}, \citenamefont {Bate},
  \citenamefont {Meraner}, \citenamefont {Sangouard}, \citenamefont {Lanyon},\
  and\ \citenamefont {Northup}}]{Krutyanskiy2022-vh}%
  \BibitemOpen
  \bibfield  {author} {\bibinfo {author} {\bibfnamefont {V.}~\bibnamefont
  {Krutyanskiy}}, \bibinfo {author} {\bibfnamefont {M.}~\bibnamefont {Galli}},
  \bibinfo {author} {\bibfnamefont {V.}~\bibnamefont {Krcmarsky}}, \bibinfo
  {author} {\bibfnamefont {S.}~\bibnamefont {Baier}}, \bibinfo {author}
  {\bibfnamefont {D.~A.}\ \bibnamefont {Fioretto}}, \bibinfo {author}
  {\bibfnamefont {Y.}~\bibnamefont {Pu}}, \bibinfo {author} {\bibfnamefont
  {A.}~\bibnamefont {Mazloom}}, \bibinfo {author} {\bibfnamefont
  {P.}~\bibnamefont {Sekatski}}, \bibinfo {author} {\bibfnamefont
  {M.}~\bibnamefont {Canteri}}, \bibinfo {author} {\bibfnamefont
  {M.}~\bibnamefont {Teller}}, \bibinfo {author} {\bibfnamefont
  {J.}~\bibnamefont {Schupp}}, \bibinfo {author} {\bibfnamefont
  {J.}~\bibnamefont {Bate}}, \bibinfo {author} {\bibfnamefont {M.}~\bibnamefont
  {Meraner}}, \bibinfo {author} {\bibfnamefont {N.}~\bibnamefont {Sangouard}},
  \bibinfo {author} {\bibfnamefont {B.~P.}\ \bibnamefont {Lanyon}},\ and\
  \bibinfo {author} {\bibfnamefont {T.~E.}\ \bibnamefont {Northup}},\
  }\bibfield  {title} {\bibinfo {title} {Entanglement of {Trapped-Ion} qubits
  separated by 230 meters},\ }\href
  {http://dx.doi.org/10.1103/PhysRevLett.130.050803} {\bibfield  {journal}
  {\bibinfo  {journal} {Phys. Rev. Lett.}\ }\textbf {\bibinfo {volume} {130}},\
  \bibinfo {pages} {050803} (\bibinfo {year} {2023})}\BibitemShut {NoStop}%
\bibitem [{\citenamefont {Bock}\ \emph {et~al.}(2018)\citenamefont {Bock},
  \citenamefont {Eich}, \citenamefont {Kucera}, \citenamefont {Kreis},
  \citenamefont {Lenhard}, \citenamefont {Becher},\ and\ \citenamefont
  {Eschner}}]{Bock2018-qx}%
  \BibitemOpen
  \bibfield  {author} {\bibinfo {author} {\bibfnamefont {M.}~\bibnamefont
  {Bock}}, \bibinfo {author} {\bibfnamefont {P.}~\bibnamefont {Eich}}, \bibinfo
  {author} {\bibfnamefont {S.}~\bibnamefont {Kucera}}, \bibinfo {author}
  {\bibfnamefont {M.}~\bibnamefont {Kreis}}, \bibinfo {author} {\bibfnamefont
  {A.}~\bibnamefont {Lenhard}}, \bibinfo {author} {\bibfnamefont
  {C.}~\bibnamefont {Becher}},\ and\ \bibinfo {author} {\bibfnamefont
  {J.}~\bibnamefont {Eschner}},\ }\bibfield  {title} {\bibinfo {title}
  {High-fidelity entanglement between a trapped ion and a telecom photon via
  quantum frequency conversion},\ }\href
  {http://dx.doi.org/10.1038/s41467-018-04341-2} {\bibfield  {journal}
  {\bibinfo  {journal} {Nat. Commun.}\ }\textbf {\bibinfo {volume} {9}},\
  \bibinfo {pages} {1998} (\bibinfo {year} {2018})}\BibitemShut {NoStop}%
\bibitem [{\citenamefont {Dudin}\ \emph {et~al.}(2010)\citenamefont {Dudin},
  \citenamefont {Radnaev}, \citenamefont {Zhao}, \citenamefont {Blumoff},
  \citenamefont {Kennedy},\ and\ \citenamefont {Kuzmich}}]{Dudin2010-ip}%
  \BibitemOpen
  \bibfield  {author} {\bibinfo {author} {\bibfnamefont {Y.~O.}\ \bibnamefont
  {Dudin}}, \bibinfo {author} {\bibfnamefont {A.~G.}\ \bibnamefont {Radnaev}},
  \bibinfo {author} {\bibfnamefont {R.}~\bibnamefont {Zhao}}, \bibinfo {author}
  {\bibfnamefont {J.~Z.}\ \bibnamefont {Blumoff}}, \bibinfo {author}
  {\bibfnamefont {T.~A.~B.}\ \bibnamefont {Kennedy}},\ and\ \bibinfo {author}
  {\bibfnamefont {A.}~\bibnamefont {Kuzmich}},\ }\bibfield  {title} {\bibinfo
  {title} {Entanglement of light-shift compensated atomic spin waves with
  telecom light},\ }\href {http://dx.doi.org/10.1103/PhysRevLett.105.260502}
  {\bibfield  {journal} {\bibinfo  {journal} {Phys. Rev. Lett.}\ }\textbf
  {\bibinfo {volume} {105}},\ \bibinfo {pages} {260502} (\bibinfo {year}
  {2010})}\BibitemShut {NoStop}%
\bibitem [{\citenamefont {van Leent}\ \emph {et~al.}(2020)\citenamefont {van
  Leent}, \citenamefont {Bock}, \citenamefont {Garthoff}, \citenamefont
  {Redeker}, \citenamefont {Zhang}, \citenamefont {Bauer}, \citenamefont
  {Rosenfeld}, \citenamefont {Becher},\ and\ \citenamefont
  {Weinfurter}}]{Van_Leent2020-ga}%
  \BibitemOpen
  \bibfield  {author} {\bibinfo {author} {\bibfnamefont {T.}~\bibnamefont {van
  Leent}}, \bibinfo {author} {\bibfnamefont {M.}~\bibnamefont {Bock}}, \bibinfo
  {author} {\bibfnamefont {R.}~\bibnamefont {Garthoff}}, \bibinfo {author}
  {\bibfnamefont {K.}~\bibnamefont {Redeker}}, \bibinfo {author} {\bibfnamefont
  {W.}~\bibnamefont {Zhang}}, \bibinfo {author} {\bibfnamefont
  {T.}~\bibnamefont {Bauer}}, \bibinfo {author} {\bibfnamefont
  {W.}~\bibnamefont {Rosenfeld}}, \bibinfo {author} {\bibfnamefont
  {C.}~\bibnamefont {Becher}},\ and\ \bibinfo {author} {\bibfnamefont
  {H.}~\bibnamefont {Weinfurter}},\ }\bibfield  {title} {\bibinfo {title}
  {{Long-Distance} distribution of {Atom-Photon} entanglement at telecom
  wavelength},\ }\href {http://dx.doi.org/10.1103/PhysRevLett.124.010510}
  {\bibfield  {journal} {\bibinfo  {journal} {Phys. Rev. Lett.}\ }\textbf
  {\bibinfo {volume} {124}},\ \bibinfo {pages} {010510} (\bibinfo {year}
  {2020})}\BibitemShut {NoStop}%
\bibitem [{\citenamefont {van Leent}\ \emph {et~al.}(2022)\citenamefont {van
  Leent}, \citenamefont {Bock}, \citenamefont {Fertig}, \citenamefont
  {Garthoff}, \citenamefont {Eppelt}, \citenamefont {Zhou}, \citenamefont
  {Malik}, \citenamefont {Seubert}, \citenamefont {Bauer}, \citenamefont
  {Rosenfeld}, \citenamefont {Zhang}, \citenamefont {Becher},\ and\
  \citenamefont {Weinfurter}}]{Van_Leent2022-yz}%
  \BibitemOpen
  \bibfield  {author} {\bibinfo {author} {\bibfnamefont {T.}~\bibnamefont {van
  Leent}}, \bibinfo {author} {\bibfnamefont {M.}~\bibnamefont {Bock}}, \bibinfo
  {author} {\bibfnamefont {F.}~\bibnamefont {Fertig}}, \bibinfo {author}
  {\bibfnamefont {R.}~\bibnamefont {Garthoff}}, \bibinfo {author}
  {\bibfnamefont {S.}~\bibnamefont {Eppelt}}, \bibinfo {author} {\bibfnamefont
  {Y.}~\bibnamefont {Zhou}}, \bibinfo {author} {\bibfnamefont {P.}~\bibnamefont
  {Malik}}, \bibinfo {author} {\bibfnamefont {M.}~\bibnamefont {Seubert}},
  \bibinfo {author} {\bibfnamefont {T.}~\bibnamefont {Bauer}}, \bibinfo
  {author} {\bibfnamefont {W.}~\bibnamefont {Rosenfeld}}, \bibinfo {author}
  {\bibfnamefont {W.}~\bibnamefont {Zhang}}, \bibinfo {author} {\bibfnamefont
  {C.}~\bibnamefont {Becher}},\ and\ \bibinfo {author} {\bibfnamefont
  {H.}~\bibnamefont {Weinfurter}},\ }\bibfield  {title} {\bibinfo {title}
  {Entangling single atoms over 33 km telecom fibre},\ }\href
  {http://dx.doi.org/10.1038/s41586-022-04764-4} {\bibfield  {journal}
  {\bibinfo  {journal} {Nature}\ }\textbf {\bibinfo {volume} {607}},\ \bibinfo
  {pages} {69} (\bibinfo {year} {2022})}\BibitemShut {NoStop}%
\bibitem [{\citenamefont {Ikuta}\ \emph {et~al.}(2018)\citenamefont {Ikuta},
  \citenamefont {Kobayashi}, \citenamefont {Kawakami}, \citenamefont {Miki},
  \citenamefont {Yabuno}, \citenamefont {Yamashita}, \citenamefont {Terai},
  \citenamefont {Koashi}, \citenamefont {Mukai}, \citenamefont {Yamamoto},\
  and\ \citenamefont {Imoto}}]{Ikuta2018-tb}%
  \BibitemOpen
  \bibfield  {author} {\bibinfo {author} {\bibfnamefont {R.}~\bibnamefont
  {Ikuta}}, \bibinfo {author} {\bibfnamefont {T.}~\bibnamefont {Kobayashi}},
  \bibinfo {author} {\bibfnamefont {T.}~\bibnamefont {Kawakami}}, \bibinfo
  {author} {\bibfnamefont {S.}~\bibnamefont {Miki}}, \bibinfo {author}
  {\bibfnamefont {M.}~\bibnamefont {Yabuno}}, \bibinfo {author} {\bibfnamefont
  {T.}~\bibnamefont {Yamashita}}, \bibinfo {author} {\bibfnamefont
  {H.}~\bibnamefont {Terai}}, \bibinfo {author} {\bibfnamefont
  {M.}~\bibnamefont {Koashi}}, \bibinfo {author} {\bibfnamefont
  {T.}~\bibnamefont {Mukai}}, \bibinfo {author} {\bibfnamefont
  {T.}~\bibnamefont {Yamamoto}},\ and\ \bibinfo {author} {\bibfnamefont
  {N.}~\bibnamefont {Imoto}},\ }\bibfield  {title} {\bibinfo {title}
  {Polarization insensitive frequency conversion for an atom-photon
  entanglement distribution via a telecom network},\ }\href
  {http://dx.doi.org/10.1038/s41467-018-04338-x} {\bibfield  {journal}
  {\bibinfo  {journal} {Nat. Commun.}\ }\textbf {\bibinfo {volume} {9}},\
  \bibinfo {pages} {1997} (\bibinfo {year} {2018})}\BibitemShut {NoStop}%
\bibitem [{\citenamefont {Borregaard}\ \emph {et~al.}(2015)\citenamefont
  {Borregaard}, \citenamefont {K{\'o}m{\'a}r}, \citenamefont {Kessler},
  \citenamefont {Lukin},\ and\ \citenamefont
  {S{\o}rensen}}]{Borregaard2015-uo}%
  \BibitemOpen
  \bibfield  {author} {\bibinfo {author} {\bibfnamefont {J.}~\bibnamefont
  {Borregaard}}, \bibinfo {author} {\bibfnamefont {P.}~\bibnamefont
  {K{\'o}m{\'a}r}}, \bibinfo {author} {\bibfnamefont {E.~M.}\ \bibnamefont
  {Kessler}}, \bibinfo {author} {\bibfnamefont {M.~D.}\ \bibnamefont {Lukin}},\
  and\ \bibinfo {author} {\bibfnamefont {A.~S.}\ \bibnamefont {S{\o}rensen}},\
  }\bibfield  {title} {\bibinfo {title} {Long-distance entanglement
  distribution using individual atoms in optical cavities},\ }\href
  {https://link.aps.org/doi/10.1103/PhysRevA.92.012307} {\bibfield  {journal}
  {\bibinfo  {journal} {Phys. Rev. A}\ }\textbf {\bibinfo {volume} {92}},\
  \bibinfo {pages} {012307} (\bibinfo {year} {2015})}\BibitemShut {NoStop}%
\bibitem [{\citenamefont {Magnard}\ \emph {et~al.}(2020)\citenamefont
  {Magnard}, \citenamefont {Storz}, \citenamefont {Kurpiers}, \citenamefont
  {Sch{\"a}r}, \citenamefont {Marxer}, \citenamefont {L{\"u}tolf},
  \citenamefont {Walter}, \citenamefont {Besse}, \citenamefont {Gabureac},
  \citenamefont {Reuer}, \citenamefont {Akin}, \citenamefont {Royer},
  \citenamefont {Blais},\ and\ \citenamefont {Wallraff}}]{Magnard2020-dc}%
  \BibitemOpen
  \bibfield  {author} {\bibinfo {author} {\bibfnamefont {P.}~\bibnamefont
  {Magnard}}, \bibinfo {author} {\bibfnamefont {S.}~\bibnamefont {Storz}},
  \bibinfo {author} {\bibfnamefont {P.}~\bibnamefont {Kurpiers}}, \bibinfo
  {author} {\bibfnamefont {J.}~\bibnamefont {Sch{\"a}r}}, \bibinfo {author}
  {\bibfnamefont {F.}~\bibnamefont {Marxer}}, \bibinfo {author} {\bibfnamefont
  {J.}~\bibnamefont {L{\"u}tolf}}, \bibinfo {author} {\bibfnamefont
  {T.}~\bibnamefont {Walter}}, \bibinfo {author} {\bibfnamefont {J.-C.}\
  \bibnamefont {Besse}}, \bibinfo {author} {\bibfnamefont {M.}~\bibnamefont
  {Gabureac}}, \bibinfo {author} {\bibfnamefont {K.}~\bibnamefont {Reuer}},
  \bibinfo {author} {\bibfnamefont {A.}~\bibnamefont {Akin}}, \bibinfo {author}
  {\bibfnamefont {B.}~\bibnamefont {Royer}}, \bibinfo {author} {\bibfnamefont
  {A.}~\bibnamefont {Blais}},\ and\ \bibinfo {author} {\bibfnamefont
  {A.}~\bibnamefont {Wallraff}},\ }\bibfield  {title} {\bibinfo {title}
  {Microwave quantum link between superconducting circuits housed in spatially
  separated cryogenic systems},\ }\href
  {http://dx.doi.org/10.1103/PhysRevLett.125.260502} {\bibfield  {journal}
  {\bibinfo  {journal} {Phys. Rev. Lett.}\ }\textbf {\bibinfo {volume} {125}},\
  \bibinfo {pages} {260502} (\bibinfo {year} {2020})}\BibitemShut {NoStop}%
\bibitem [{\citenamefont {Krastanov}\ \emph {et~al.}(2021)\citenamefont
  {Krastanov}, \citenamefont {Raniwala}, \citenamefont {Holzgrafe},
  \citenamefont {Jacobs}, \citenamefont {Lon{\v c}ar}, \citenamefont {Reagor},\
  and\ \citenamefont {Englund}}]{Krastanov2021-pq}%
  \BibitemOpen
  \bibfield  {author} {\bibinfo {author} {\bibfnamefont {S.}~\bibnamefont
  {Krastanov}}, \bibinfo {author} {\bibfnamefont {H.}~\bibnamefont {Raniwala}},
  \bibinfo {author} {\bibfnamefont {J.}~\bibnamefont {Holzgrafe}}, \bibinfo
  {author} {\bibfnamefont {K.}~\bibnamefont {Jacobs}}, \bibinfo {author}
  {\bibfnamefont {M.}~\bibnamefont {Lon{\v c}ar}}, \bibinfo {author}
  {\bibfnamefont {M.~J.}\ \bibnamefont {Reagor}},\ and\ \bibinfo {author}
  {\bibfnamefont {D.~R.}\ \bibnamefont {Englund}},\ }\bibfield  {title}
  {\bibinfo {title} {Optically heralded entanglement of superconducting systems
  in quantum networks},\ }\href
  {http://dx.doi.org/10.1103/PhysRevLett.127.040503} {\bibfield  {journal}
  {\bibinfo  {journal} {Phys. Rev. Lett.}\ }\textbf {\bibinfo {volume} {127}},\
  \bibinfo {pages} {040503} (\bibinfo {year} {2021})}\BibitemShut {NoStop}%
\bibitem [{\citenamefont {Jones}\ \emph {et~al.}(2013)\citenamefont {Jones},
  \citenamefont {De~Greve},\ and\ \citenamefont {Yamamoto}}]{Jones2013-wh}%
  \BibitemOpen
  \bibfield  {author} {\bibinfo {author} {\bibfnamefont {C.}~\bibnamefont
  {Jones}}, \bibinfo {author} {\bibfnamefont {K.}~\bibnamefont {De~Greve}},\
  and\ \bibinfo {author} {\bibfnamefont {Y.}~\bibnamefont {Yamamoto}},\
  }\bibfield  {title} {\bibinfo {title} {A high-speed optical link to entangle
  quantum dots},\ }\href {https://arxiv.org/abs/1310.4609} {\  (\bibinfo {year}
  {2013})},\ \Eprint {https://arxiv.org/abs/1310.4609} {arXiv:1310.4609
  [quant-ph]} \BibitemShut {NoStop}%
\bibitem [{\citenamefont {Jones}\ \emph {et~al.}(2016)\citenamefont {Jones},
  \citenamefont {Kim}, \citenamefont {Rakher}, \citenamefont {Kwiat},\ and\
  \citenamefont {Ladd}}]{Jones2016-mg}%
  \BibitemOpen
  \bibfield  {author} {\bibinfo {author} {\bibfnamefont {C.}~\bibnamefont
  {Jones}}, \bibinfo {author} {\bibfnamefont {D.}~\bibnamefont {Kim}}, \bibinfo
  {author} {\bibfnamefont {M.~T.}\ \bibnamefont {Rakher}}, \bibinfo {author}
  {\bibfnamefont {P.~G.}\ \bibnamefont {Kwiat}},\ and\ \bibinfo {author}
  {\bibfnamefont {T.~D.}\ \bibnamefont {Ladd}},\ }\bibfield  {title} {\bibinfo
  {title} {Design and analysis of communication protocols for quantum repeater
  networks},\ }\href
  {https://iopscience.iop.org/article/10.1088/1367-2630/18/8/083015} {\bibfield
   {journal} {\bibinfo  {journal} {New J. Phys.}\ }\textbf {\bibinfo {volume}
  {18}},\ \bibinfo {pages} {083015} (\bibinfo {year} {2016})}\BibitemShut
  {NoStop}%
\bibitem [{\citenamefont {Hermans}\ \emph {et~al.}(2023)\citenamefont
  {Hermans}, \citenamefont {Pompili}, \citenamefont {Dos Santos~Martins},
  \citenamefont {R-P~Montblanch}, \citenamefont {Beukers}, \citenamefont
  {Baier}, \citenamefont {Borregaard},\ and\ \citenamefont
  {Hanson}}]{Hermans2023-kf}%
  \BibitemOpen
  \bibfield  {author} {\bibinfo {author} {\bibfnamefont {S.~L.~N.}\
  \bibnamefont {Hermans}}, \bibinfo {author} {\bibfnamefont {M.}~\bibnamefont
  {Pompili}}, \bibinfo {author} {\bibfnamefont {L.}~\bibnamefont {Dos
  Santos~Martins}}, \bibinfo {author} {\bibfnamefont {A.}~\bibnamefont
  {R-P~Montblanch}}, \bibinfo {author} {\bibfnamefont {H.~K.~C.}\ \bibnamefont
  {Beukers}}, \bibinfo {author} {\bibfnamefont {S.}~\bibnamefont {Baier}},
  \bibinfo {author} {\bibfnamefont {J.}~\bibnamefont {Borregaard}},\ and\
  \bibinfo {author} {\bibfnamefont {R.}~\bibnamefont {Hanson}},\ }\bibfield
  {title} {\bibinfo {title} {Entangling remote qubits using the single-photon
  protocol: an in-depth theoretical and experimental study},\ }\href
  {https://iopscience.iop.org/article/10.1088/1367-2630/acb004/meta} {\bibfield
   {journal} {\bibinfo  {journal} {New J. Phys.}\ }\textbf {\bibinfo {volume}
  {25}},\ \bibinfo {pages} {013011} (\bibinfo {year} {2023})}\BibitemShut
  {NoStop}%
\bibitem [{\citenamefont {Goodenough}\ \emph {et~al.}(2021)\citenamefont
  {Goodenough}, \citenamefont {Elkouss},\ and\ \citenamefont
  {Wehner}}]{Goodenough2021-nt}%
  \BibitemOpen
  \bibfield  {author} {\bibinfo {author} {\bibfnamefont {K.}~\bibnamefont
  {Goodenough}}, \bibinfo {author} {\bibfnamefont {D.}~\bibnamefont
  {Elkouss}},\ and\ \bibinfo {author} {\bibfnamefont {S.}~\bibnamefont
  {Wehner}},\ }\bibfield  {title} {\bibinfo {title} {Optimizing repeater
  schemes for the quantum internet},\ }\href
  {https://link.aps.org/doi/10.1103/PhysRevA.103.032610} {\bibfield  {journal}
  {\bibinfo  {journal} {Phys. Rev. A}\ }\textbf {\bibinfo {volume} {103}},\
  \bibinfo {pages} {032610} (\bibinfo {year} {2021})}\BibitemShut {NoStop}%
\bibitem [{\citenamefont {Wein}\ \emph {et~al.}(2020)\citenamefont {Wein},
  \citenamefont {Ji}, \citenamefont {Wu}, \citenamefont {Kimiaee~Asadi},
  \citenamefont {Ghobadi},\ and\ \citenamefont {Simon}}]{Wein2020-vp}%
  \BibitemOpen
  \bibfield  {author} {\bibinfo {author} {\bibfnamefont {S.~C.}\ \bibnamefont
  {Wein}}, \bibinfo {author} {\bibfnamefont {J.-W.}\ \bibnamefont {Ji}},
  \bibinfo {author} {\bibfnamefont {Y.-F.}\ \bibnamefont {Wu}}, \bibinfo
  {author} {\bibfnamefont {F.}~\bibnamefont {Kimiaee~Asadi}}, \bibinfo {author}
  {\bibfnamefont {R.}~\bibnamefont {Ghobadi}},\ and\ \bibinfo {author}
  {\bibfnamefont {C.}~\bibnamefont {Simon}},\ }\bibfield  {title} {\bibinfo
  {title} {Analyzing photon-count heralded entanglement generation between
  solid-state spin qubits by decomposing the master-equation dynamics},\ }\href
  {https://journals.aps.org/pra/abstract/10.1103/PhysRevA.102.033701}
  {\bibfield  {journal} {\bibinfo  {journal} {Phys. Rev. A}\ }\textbf {\bibinfo
  {volume} {102}},\ \bibinfo {pages} {033701} (\bibinfo {year}
  {2020})}\BibitemShut {NoStop}%
\bibitem [{Note1()}]{Note1}%
  \BibitemOpen
  \bibinfo {note} {The word metric is used here without any mathematical
  rigor.}\BibitemShut {Stop}%
\bibitem [{\citenamefont {Pirandola}\ \emph {et~al.}(2017)\citenamefont
  {Pirandola}, \citenamefont {Laurenza}, \citenamefont {Ottaviani},\ and\
  \citenamefont {Banchi}}]{Pirandola2017-qr}%
  \BibitemOpen
  \bibfield  {author} {\bibinfo {author} {\bibfnamefont {S.}~\bibnamefont
  {Pirandola}}, \bibinfo {author} {\bibfnamefont {R.}~\bibnamefont {Laurenza}},
  \bibinfo {author} {\bibfnamefont {C.}~\bibnamefont {Ottaviani}},\ and\
  \bibinfo {author} {\bibfnamefont {L.}~\bibnamefont {Banchi}},\ }\bibfield
  {title} {\bibinfo {title} {Fundamental limits of repeaterless quantum
  communications},\ }\href {https://www.nature.com/articles/ncomms15043}
  {\bibfield  {journal} {\bibinfo  {journal} {Nat. Commun.}\ }\textbf {\bibinfo
  {volume} {8}},\ \bibinfo {pages} {15043} (\bibinfo {year}
  {2017})}\BibitemShut {NoStop}%
\bibitem [{\citenamefont {Calsamiglia}\ and\ \citenamefont
  {L{\"u}tkenhaus}(2001)}]{Calsamiglia2001-vp}%
  \BibitemOpen
  \bibfield  {author} {\bibinfo {author} {\bibfnamefont {J.}~\bibnamefont
  {Calsamiglia}}\ and\ \bibinfo {author} {\bibfnamefont {N.}~\bibnamefont
  {L{\"u}tkenhaus}},\ }\bibfield  {title} {\bibinfo {title} {Maximum efficiency
  of a linear-optical bell-state analyzer},\ }\href
  {https://link.springer.com/article/10.1007/s003400000484} {\bibfield
  {journal} {\bibinfo  {journal} {Appl. Phys. B}\ }\textbf {\bibinfo {volume}
  {72}},\ \bibinfo {pages} {67} (\bibinfo {year} {2001})}\BibitemShut {NoStop}%
\bibitem [{\citenamefont {Deutsch}\ \emph {et~al.}(1996)\citenamefont
  {Deutsch}, \citenamefont {Ekert}, \citenamefont {Jozsa}, \citenamefont
  {Macchiavello}, \citenamefont {Popescu},\ and\ \citenamefont
  {Sanpera}}]{Deutsch1996-sb}%
  \BibitemOpen
  \bibfield  {author} {\bibinfo {author} {\bibfnamefont {D.}~\bibnamefont
  {Deutsch}}, \bibinfo {author} {\bibfnamefont {A.}~\bibnamefont {Ekert}},
  \bibinfo {author} {\bibfnamefont {R.}~\bibnamefont {Jozsa}}, \bibinfo
  {author} {\bibfnamefont {C.}~\bibnamefont {Macchiavello}}, \bibinfo {author}
  {\bibfnamefont {S.}~\bibnamefont {Popescu}},\ and\ \bibinfo {author}
  {\bibfnamefont {A.}~\bibnamefont {Sanpera}},\ }\bibfield  {title} {\bibinfo
  {title} {Quantum privacy amplification and the security of quantum
  cryptography over noisy channels},\ }\href
  {https://journals.aps.org/prl/abstract/10.1103/PhysRevLett.77.2818}
  {\bibfield  {journal} {\bibinfo  {journal} {Phys. Rev. Lett.}\ }\textbf
  {\bibinfo {volume} {77}},\ \bibinfo {pages} {2818} (\bibinfo {year}
  {1996})}\BibitemShut {NoStop}%
\bibitem [{\citenamefont {Rozpedek}\ \emph {et~al.}(2018)\citenamefont
  {Rozpedek}, \citenamefont {Schiet}, \citenamefont {Thinh}, \citenamefont
  {Elkouss}, \citenamefont {Doherty},\ and\ \citenamefont
  {Wehner}}]{Rozpedek2018-ll}%
  \BibitemOpen
  \bibfield  {author} {\bibinfo {author} {\bibfnamefont {F.}~\bibnamefont
  {Rozpedek}}, \bibinfo {author} {\bibfnamefont {T.}~\bibnamefont {Schiet}},
  \bibinfo {author} {\bibfnamefont {L.~P.}\ \bibnamefont {Thinh}}, \bibinfo
  {author} {\bibfnamefont {D.}~\bibnamefont {Elkouss}}, \bibinfo {author}
  {\bibfnamefont {A.~C.}\ \bibnamefont {Doherty}},\ and\ \bibinfo {author}
  {\bibfnamefont {S.}~\bibnamefont {Wehner}},\ }\bibfield  {title} {\bibinfo
  {title} {Optimizing practical entanglement distillation},\ }\href
  {https://link.aps.org/doi/10.1103/PhysRevA.97.062333} {\bibfield  {journal}
  {\bibinfo  {journal} {Phys. Rev. A}\ }\textbf {\bibinfo {volume} {97}},\
  \bibinfo {pages} {062333} (\bibinfo {year} {2018})}\BibitemShut {NoStop}%
\bibitem [{\citenamefont {Andrews}\ and\ \citenamefont
  {Phillips}(2005)}]{Andrews2005-zi}%
  \BibitemOpen
  \bibfield  {author} {\bibinfo {author} {\bibfnamefont {L.~C.}\ \bibnamefont
  {Andrews}}\ and\ \bibinfo {author} {\bibfnamefont {R.~L.}\ \bibnamefont
  {Phillips}},\ }\href {https://spie.org/Publications/Book/626196?SSO=1} {\emph
  {\bibinfo {title} {Laser Beam Propagation Through Random Media}}}\ (\bibinfo
  {publisher} {SPIE Digital Library},\ \bibinfo {year} {2005})\BibitemShut
  {NoStop}%
\bibitem [{\citenamefont {Belmonte}\ and\ \citenamefont
  {Kahn}(2009)}]{Belmonte2009-hb}%
  \BibitemOpen
  \bibfield  {author} {\bibinfo {author} {\bibfnamefont {A.}~\bibnamefont
  {Belmonte}}\ and\ \bibinfo {author} {\bibfnamefont {J.~M.}\ \bibnamefont
  {Kahn}},\ }\bibfield  {title} {\bibinfo {title} {Capacity of coherent
  free-space optical links using diversity-combining techniques},\ }\href
  {http://dx.doi.org/10.1364/oe.17.012601} {\bibfield  {journal} {\bibinfo
  {journal} {Opt. Express}\ }\textbf {\bibinfo {volume} {17}},\ \bibinfo
  {pages} {12601} (\bibinfo {year} {2009})}\BibitemShut {NoStop}%
\bibitem [{\citenamefont {Noll}(1976)}]{Noll1976-aw}%
  \BibitemOpen
  \bibfield  {author} {\bibinfo {author} {\bibfnamefont {R.~J.}\ \bibnamefont
  {Noll}},\ }\bibfield  {title} {\bibinfo {title} {Zernike polynomials and
  atmospheric turbulence},\ }\href
  {https://opg.optica.org/abstract.cfm?uri=josa-66-3-207} {\bibfield  {journal}
  {\bibinfo  {journal} {J. Opt. Soc. Am., JOSA}\ }\textbf {\bibinfo {volume}
  {66}},\ \bibinfo {pages} {207} (\bibinfo {year} {1976})}\BibitemShut
  {NoStop}%
\bibitem [{\citenamefont {Bennett}\ \emph {et~al.}(1996)\citenamefont
  {Bennett}, \citenamefont {DiVincenzo}, \citenamefont {Smolin},\ and\
  \citenamefont {Wootters}}]{Bennett1996-vn}%
  \BibitemOpen
  \bibfield  {author} {\bibinfo {author} {\bibfnamefont {C.~H.}\ \bibnamefont
  {Bennett}}, \bibinfo {author} {\bibfnamefont {D.~P.}\ \bibnamefont
  {DiVincenzo}}, \bibinfo {author} {\bibfnamefont {J.~A.}\ \bibnamefont
  {Smolin}},\ and\ \bibinfo {author} {\bibfnamefont {W.~K.}\ \bibnamefont
  {Wootters}},\ }\bibfield  {title} {\bibinfo {title} {Mixed-state entanglement
  and quantum error correction},\ }\href
  {https://journals.aps.org/pra/abstract/10.1103/PhysRevA.54.3824} {\bibfield
  {journal} {\bibinfo  {journal} {Phys. Rev. A}\ }\textbf {\bibinfo {volume}
  {54}},\ \bibinfo {pages} {3824} (\bibinfo {year} {1996})}\BibitemShut
  {NoStop}%
\end{thebibliography}%

\end{document}